\documentclass[fleqn, twocolumn, notitlepage, amsmath,amssymb,superscriptaddress]{revtex4}
\usepackage{epsfig}
\usepackage{amsmath}
\usepackage{graphicx}
\usepackage{color}
\usepackage{amsfonts,amssymb}
\usepackage[english]{babel}
\usepackage{hyperref}
\usepackage{lipsum}
\usepackage{graphics}
\usepackage{array}
\usepackage{tabu}
\usepackage{bm}
\usepackage{float}
\usepackage{setspace}

\newcommand{\ct}{\cite}

\newcommand{\be}{\begin{equation}}
\newcommand{\ee}{\end{equation}}
\newcommand{\ba}{\begin{eqnarray}}
\newcommand{\ea}{\end{eqnarray}}

\newcommand{\non}{\nonumber}
\newcommand{\bra}[1]{\langle #1|}
\newcommand{\ket}[1]{|#1\rangle}

\begin{document}

\title{Anti-Zeno quantum advantage in fast-driven heat machines}

\author{Victor Mukherjee}
\email{mukherjeev@iiserbpr.ac.in}
\affiliation{International Center of Quantum Artificial Intelligence for Science and Technology (QuArtist) \\ and Department of Physics, Shanghai University, 200444 Shanghai, China}
\affiliation{Department of Chemical and Biological Physics, Weizmann Institute of Science, Rehovot 7610001, Israel}
\affiliation{Department of Physical Sciences, IISER Berhampur, Berhampur 760010, India}

\author{Abraham G. Kofman}
\email{kofmana@gmail.com}
\affiliation{International Center of Quantum Artificial Intelligence for Science and Technology (QuArtist) \\ and Department of Physics, Shanghai University, 200444 Shanghai, China}
\affiliation{Department of Chemical and Biological Physics, Weizmann Institute of Science, Rehovot 7610001, Israel}

\author{Gershon Kurizki}
\affiliation{Department of Chemical and Biological Physics, Weizmann Institute of Science, Rehovot 7610001, Israel}

\begin{abstract}
Developing quantum machines which can outperform their classical counterparts, thereby achieving quantum supremacy or quantum advantage, is a major aim of the current research on quantum 
thermodynamics and quantum technologies.
Here we show that a fast-modulated cyclic quantum heat machine operating in the non-Markovian regime can lead to significant heat-current and power boosts induced by 
the anti-Zeno effect. Such  boosts signify a 
quantum advantage over almost all heat-machines proposed thus far that operate in the conventional Markovian regime, where the quantumness of the system-bath interaction plays no role. The
present effect owes its origin to the
time-energy uncertainty relation in quantum mechanics, which may result in enhanced system-bath energy exchange for modulation periods shorter than the bath correlation-time.

\end{abstract}

\maketitle

\section*{Introduction}

The non-equilibrium thermodynamic description of heat machines consisting of quantum systems coupled to heat baths is almost exclusively based on the Markovian approximation \cite{breuer02, rivas12open}.
This approximation allows for monotonic convergence 
of the system-state to thermal equilibrium  with its environment (bath) and yields a universal bound on entropy change (production) in the system \cite{spohn78entropy}. Yet, the 
Markovian approximation is not required for the derivation of the Carnot bound on the efficiency of a cyclic two-bath heat engine (HE): this bound follows  from the second 
law of thermodynamics, 
under the condition of zero entropy change  over a cycle by the working fluid (WF), in both classical and quantum scenarios. In general, the question whether non-Markovianity is an asset remains open, 
although several works have ventured into the non-Markovian domain \cite{mukherjee15, uzdin16quantum, pezzutto18an, thomas18thermodynamics, nahar18preparations, abiuso19non}.  By contrast,
it has been suggested that quantum resources, such as a bath consisting of coherently superposed atoms \cite{scully03extracting}, or a squeezed thermal
bath \cite{rossnage14nanoscale, klaers17squeezed, niedenzu18quantum}, 
may raise the efficiency bound of the machine. The mechanisms that can cause such a raise include either a conversion of atomic coherence and entanglement in the 
bath into WF
heatup \cite{scully03extracting, abah14efficiency, dag18temperature}, or the ability of a squeezed bath to exchange ergotropy \cite{rossnage14nanoscale, niedenzu16on, klaers17squeezed, niedenzu18quantum}
(alias non-passivity or work-capacity \cite{pusz78, lenard78, klimovsky15}) with the WF, which is incompatible with a standard HE. However, 
neither of these mechanisms  is exclusively quantum; both may have classical counterparts \cite{ghosh18are}. Likewise, quantum coherent or squeezed driving of the system acting as a WF 
or a piston \cite{ghosh17catalysis} may boost the power output of the machine depending on the ergotropy of the system state, but not on its non-classicality \cite{niedenzu18quantum}.

Finding quantum advantages in machine performance relative to their classical counterparts has been one of the major aims of research in the field of quantum technology
in general \cite{kurizki15quantum, harrow17quantum, boixo18characterizing}, and particularly in thermodynamics of quantum systems \cite{ghosh2018thermodynamic}. Overall, the
foregoing research leads to the conclusion that conventional thermodynamic description of cyclic machines based on  a (two-level, multilevel or harmonic oscillator) quantum system 
in arbitrary two-bath settings may not be the arena for a distinct quantum advantage  in machine performance \cite{ghosh18are}. 
An exception should be made for multiple identical machines that exhibit collective, quantum-entangled
features \cite{niedenzu18cooperative, campo16quantum}).

Here we show that quantum advantage is in fact achievable in a quantum heat machine (QHM), whether a heat engine or a refrigerator,  whose
energy-level gap is modulated faster than what is allowed by the Markov approximation.  To this end, we invoke
methods of 
quantum system-control via frequent coherent (e.g. phase-flipping or level-modulating) operations \cite{kofman01universal, kofman04unified} as well as their incoherent counterparts (e.g. 
projective measurements 
or noise-induced dephasing) \cite{kofman00acceleration, erez08, gordon09cooling, gordon10equilibration, alvarez10zeno}. Such control has previously been shown, 
both theoretically\cite{kofman00acceleration, gordon07universal, erez08, clausen10bath} and 
experimentally \cite{alvarez10zeno, almog11direct}, to yield 
non-Markovian  dynamics that 
conforms to one of two universal paradigms: i) quantum Zeno dynamics (QZD) whereby the bath effects on the system are drastically suppressed or slowed down;
ii) anti-Zeno dynamics (AZD) that implies the opposite, i.e., enhancement or speedup of the system-bath energy exchange \cite{kofman00acceleration, erez08, rao11from}. It has been previously shown that 
QZD 
leads to the heating of both the system and the bath 
at the expense of the system-bath correlation energy \cite{klimovsky13work}, whereas AZD may lead to alternating cooling or heating of the system at the expense 
of the bath or 
vice-versa \cite{kofman00acceleration, erez08}.
In our present analysis of cyclic heat 
machines based on quantum systems, we show that analogous effects can drastically modify the power output, without affecting their Carnot efficiency bound. AZD is shown to bring about a drastic 
power boost, thereby manifesting genuine quantum advantage, as it stems from the time-energy uncertainty relation of quantum mechanics. 

\section*{Results}

\begin{figure}[t]
\begin{center}
\includegraphics[width = 0.85\columnwidth]{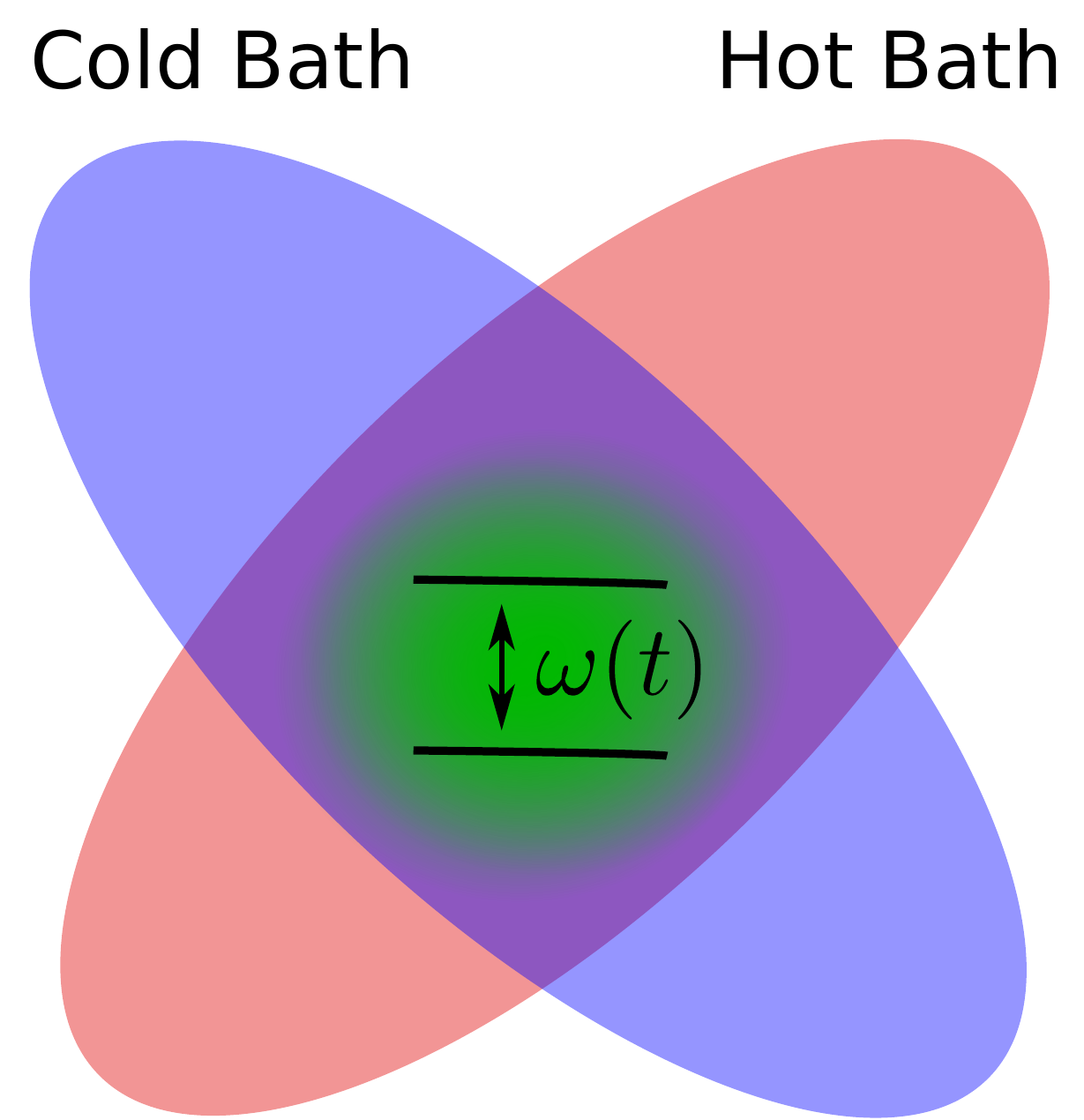}
\end{center}
\caption{{\bf A two-level system thermal machine:} Schematic setup showing a two-level system with periodically modulated level distance $\omega(t)$ as the working fluid (WF) in a thermal machine wherein the WF is simultaneously coupled to hot and cold baths with non-overlapping spectra. Possible realizations include a driven-atom WF coupled to filtered heat baths in a cavity or a driven impurity coupled to spectrally distinct phonon baths in a solid structure \cite{klimovsky13minimal, ghosh18we}.}
\label{schem}
\end{figure}

{\bf Model.}~We consider a quantum system $\mathcal{S}$ that plays the role of a working fluid (WF)  in a quantum thermal machine, wherein it is
simultaneously coupled to cold and
hot thermal baths.
The system is periodically driven  or perturbed with time period $\tau_{\rm S} = 2\pi/\Delta_{\rm S}$ by the time-dependent Hamiltonian $\hat{H}_{\rm S}(t)$:
\ba
\hat{H}_{\rm S}(t + \tau_{\rm S}) = \hat{H}_{\rm S}(t).
\label{hper}
\ea
In order to have frictionless dynamics at all times, we choose $\hat{H}_{\rm S}(t)$ to be diagonal in the energy basis of $\mathcal{S}$, such that.
\ba
\left[\hat{H}_{\rm S}(t), \hat{H}_{\rm S}(t^{\prime}) \right] &=& 0 ~~~ \forall~t,t^{\prime}.
\label{hamil}
\ea
The system interacts simultaneously with the independent cold (c) and hot (h) baths via 
\ba
\hat{H}_{\rm I} = \sum_{j = {\rm c,h}}  \hat{S}\otimes \hat{B}_j,
\label{hamilint}
\ea
where the  bath operators $\hat{B}_{\rm c}$ and $\hat{B}_{\rm h}$ commute: $\left[\hat{B}_{\rm c}, \hat{B}_{\rm h} \right] = 0$, and
$\hat{S}$ is a system operator. For example, for a two level system, $\hat{S} = \hat{\sigma}_x$, while  $\hat{S} =  \hat{X}$
for a harmonic oscillator, in standard notations. We do  not
invoke the rotating wave approximation in the system-bath interaction Hamiltonian Eq. \eqref{hamilint}. 
As in the minimal continuous quantum heat machine \cite{klimovsky13minimal}, or its multilevel extensions \cite{mukherjee16speed},
we require the two baths to have non-overlapping spectra, e.g.,
super-Ohmic spectra with 
distinct upper cut-off frequencies (see Fig. \ref{schem}). This requirement allows $\mathcal{S}$ to effectively couple intermittently to one or
the other bath during the modulation period $\tau_{\rm S}$, 
without changing the interaction Hamiltonian 
to either bath. 

{\bf From Markovian to non-Markovian dynamics.}~In what follows we assume weak system-bath coupling, consistent with the Born (but not necessarily the Markov) approximation. 
Our goal is to examine the dynamics as we transit from Markovian to non-Markovian time scales, and the ensuing change of the QHM performance as the period 
duration $\tau_{\rm S}$ is decreased. To this end, we have adopted the methodology previously derived in Refs. \cite{kofman01universal, kofman04unified, shahmoon13, kosloff13quantum},
to account for the periodicity of $\hat{H}_{\rm S}(t)$, 
by resorting to a Floquet expansion of the Liouville operator in the harmonics of $\Delta_{\rm S} = 2\pi/\tau_{\rm S}$ \cite{szczygielski13, alicki14quantum, klimovsky13minimal}. 
As explained below, we focus on system-bath coupling durations 
$\tau_{\rm C} = n\tau_{\rm S}$ of the order of a few modulation periods, where $n > 1$ denotes the number of periods. 
The time-scales of importance are the modulation time period $\tau_{\rm S}$, the system-bath coupling  duration $\tau_{\rm C}$, the bath correlation-time $\tau_{\rm B}$
and the thermalization time $\tau_{\rm th} \sim \gamma_0^{-1}$, where $\gamma_0$ is the  system-bath 
coupling strength. 
We consider $n \gg 1$ such that $\tau_{\rm C} \gg \tau_{\rm S},~\left(\omega + q\Delta_{\rm S} \right)^{-1}$, where $\omega$ denotes
the transition frequencies of the system $\mathcal{S}$, and $q$ is an integer 
(see Methods ``Floquet Analysis of the non-Markovian
Master Equation''). This allows us to implement the secular approximation, thereby averaging over the fast-rotating 
terms in the dynamics. 
In the limit of slow modulation, i.e, $\tau_{\rm S} \gg \tau_{\rm B}$, we have $\tau_{\rm C} \gg \tau_{\rm B}$, which allows us
to perform the Born, Markov and secular approximations, and eventually arrive at a time-independent
Markovian master 
equation for $\tau_{\rm C} \gg \tau_{\rm S},~\omega^{-1},~\tau_{\rm B}$ (see Methods ``Floquet Analysis of the non-Markovian
Master Equation'').

On the other hand, in the regime of fast modulation $\tau_{\rm S} \ll \tau_{\rm B}$, the Markov approximation becomes inapplicable for coupling durations
$\tau_{\rm C} = n\tau_{\rm S} \lesssim \tau_{\rm B}$.
This gives rise to the fast-modulation form of the master equation (see Methods. ``Floquet Analysis of the non-Markovian
Master Equation'' and ``Non-Markovian dynamics of a driven two-level system in a dissipative bath''): 
\ba
\dot{\rho}_{\rm S}(t) &=& \sum_{j = {\rm h, c}} \mathcal{L}_j\left[\rho_{\rm S}(t) \right] \non\\
&=& \sum_{j, \omega}\tilde{\mathcal{I}}_{j}(\omega,t)  \mathcal{D}_{j,\omega}\left[\rho_{\rm S}(t) \right] + \text{h.c.};\non\\
\tilde{\mathcal{I}}_j(\omega,t) &\equiv& \int_{-\infty}^{\infty} d\nu G_j(\nu)\Big[\frac{\sin\left[\left(\nu - \omega \right)t\right]}{\nu - \omega} \non\\
&\pm& i \left(\frac{\cos\left[\left(\nu - \omega \right)t\right] - 1}{\nu - \omega}\right)\Big]
\label{rwaTsinc}
\ea
For simplicity, unless otherwise stated, we consider $\hbar = k_{\rm B} = 1$. Here, for any
modulation period $\tau_{\rm S}$, the generalized
Liouville operators $\mathcal{L}_j$ of the two 
baths act additively on the reduced density matrix $\rho_{\rm S}(t)$ of $\mathcal{S}$, 
generated by the $\omega$-spectral 
components of the Lindblad dissipators $\mathcal{D}_{j,\omega}$ (see below) for the $j = {\rm c, h}$ bath acting on $\rho_{\rm S}(t)$. For a two-level system,
or an oscillator, $\mathcal{D}$ does not depend on 
$\omega$ \cite{breuer02}. For $\rho_{\rm S}(t)$ that is diagonal in the energy basis, which we consider below,
the dynamics is dictated by the coefficients 
$\mathcal{I}_{j}(\omega,t) \equiv {\rm Re}\left[\tilde{\mathcal{I}}_j(\omega,t)\right]$ in Eq. \eqref{rwaTsinc}, which 
express the convolution of the $j$-th bath spectral response function $G_{j}(\nu)$
that has spectral width $\sim \Gamma_{\rm B} \sim 1/\tau_{\rm B}$,
with the sinc function, imposed by the time-energy uncertainty 
relation for finite times (see Methods ``Non-Markovian dynamics of a driven
two-level system in a dissipative bath''). 

Our main contention is that
overlap between the sinc function and $G_j(\nu)$ at $t \sim \tau_{\rm C}  \lesssim \tau_{\rm B}$
may lead to the anti-Zeno effect, i.e., to remarkable enhancement in the convolution $\mathcal{I}_{j}(\omega,t)$, and, correspondingly, in the
heat currents and power. One can stay in this regime of enhanced performance over many cycles, by running the QHM in the following two-stroke non-Markovian cycles:\\
i) Stroke 1: we run the QHM by keeping the WF (system) and the baths coupled over $n$ modulation periods, from time 
$t = 0$ to $t = n\tau_{\rm S} = \tau_{\rm C} \lesssim \tau_{\rm B}$~($n \gg 1$, $\tau_{\rm S} \ll \tau_{\rm B}$).  The $n$ modulation periods of the WF are equivalent to $n$ cycles of 
continuous heat machines studied earlier, which have been shown to exploit spectral separation of the hot and cold baths for the extraction of work \cite{klimovsky13minimal, alicki14quantum}, or 
 refrigeration \cite{klimovsky15, kolar12}, in the Markovian regime (see Eq. \eqref{specsep}). By contrast, in the non-Markovian domain a modulation period is not a cycle, since the time-dependent heat currents 
 and the WF state are not necessarily reset to their initial values at the beginning of each modulation period (see below).\\
ii) Stroke 2: In order to reset the WF state and the heat currents to their initial ($t = 0$) values in the non-Markovian regime, we have to add another stroke: At $t = n\tau_{\rm S} = \tau_{\rm C} $, we decouple the 
WF from the hot and cold baths. One needs to keep the WF and the thermal baths uncoupled (non-interacting) 
for a time-interval $\bar{t} \gtrsim \tau_{\rm B}$, so as to 
eliminate all the transient memory effects \cite{rao11from}. \\
After this decoupling period, we recouple the WF to the hot and cold thermal baths and continue to 
drive the WF with the periodically 
modulated Hamiltonian Eq. \eqref{hper}. Thus the setup is initialized
\begin{figure}[t]
\begin{center}
\includegraphics[width= 0.95\columnwidth, angle = 0]{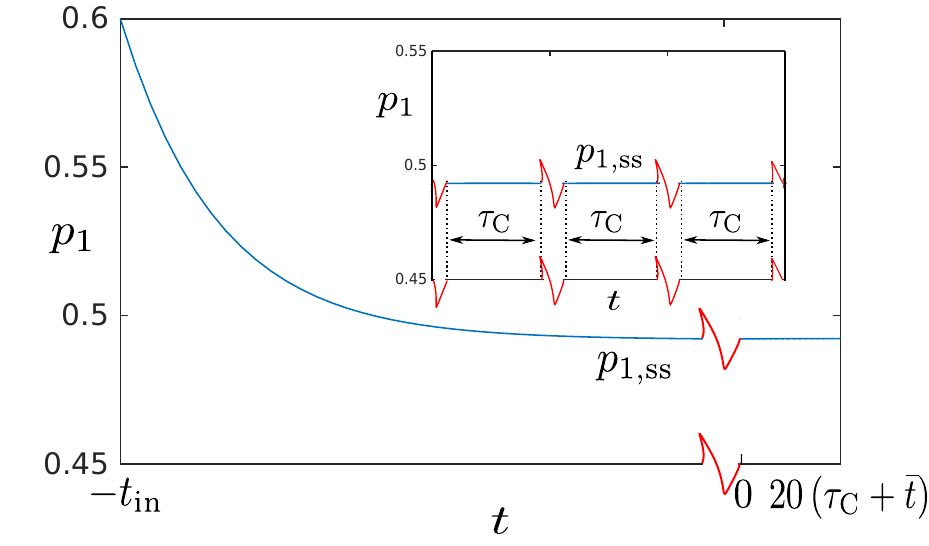}
\end{center}
\caption {{\bf Time-evolution:} Time-evolution of the $\ket{1}$-state probability $p_1(t)$ of a two-level system working fluid (WF). The WF is first connected to the 
hot and cold baths, whose quasi-Lorentzian spectral functions are given by Eq. \eqref{ghgceq}, at a negative time $-t_{in}$ $(t_{in} \gg \tau_{\rm th})$, under the initial condition
$p_1(t = -t_{\rm in}) = 0.6$, and reaches the steady-state value $p_{1, {\rm ss}}$ at 
$t + t_{\rm in} \gg \tau_{\rm th}$. 
The WF is decoupled from the hot and cold thermal baths at a time $-\bar{t} \lesssim -\tau_{\rm B} < 0$ after reaching the steady-state, and then recoupled 
again to the two baths
at time $t = 0$, such that the WF is non-interacting with the hot and cold thermal baths for the time interval $-\bar{t} \leq t <0$, shown by the red break-line. 
The quantum heat machine is operated in the anti-Zeno dynamics (AZD) regime for $t \geq 0$, wherein it is decoupled 
from and recoupled to the thermal baths after every AZD cycle, for coupling time duration 
$\tau_{\rm C} = n\tau_{\rm S}$. The probability  $p_1$ remains unchanged at
the steady-state value, even after multiple  AZD cycles. Inset: Same as the main plot, zoomed in for three consecutive AZD cycles. The WF is non-interacting with the thermal 
baths for time intervals $\bar{t} \gtrsim \tau_{\rm B}$ between two consecutive AZD cycles, shown by the
red break lines. Here (see Eqs. \eqref{modtls} - \eqref{wstate}) $\lambda = 0.2, \omega_0 = 20, \Delta_{\rm S} = 10, n = 10, \beta_{\rm h} = 0.0005, \beta_{\rm c} = 0.005$, and we consider quasi-Lorentzian bath spectral functions 
Eq. \eqref{ghgceq} with $\gamma_0 = 1, \Gamma_{\rm B} = 0.2,  \delta_{\rm h} = \delta_{\rm c} = 1, \alpha = 1.$}
\label{dyn}
\end{figure}
after time $\tau_{\rm C} + \bar{t}$, provided we choose $n$ to be such that $\rho_{\rm S}(\tau_{\rm C} + \bar{t}) = \rho_{\rm S}(0)$, so as to close the steady-state cycle after $n$ modulation periods, with the WF 
returning to its state at start of the cycle (see Fig. \ref{dyn} and Sec. ``A minimal quantum thermal machine beyond Markovianity''). The QHM may then run indefinitely in the non-Markovian cyclic regime.

By contrast, in the limit of long WF-baths coupling duration $\tau_{\rm C} = n\tau_{\rm S} \gg \tau_{\rm B}$, the ${\rm sinc}$ functions  take the form of  delta functions, and therefore,  as expected, 
the integral Eq. \eqref{rwaTsinc} reduces to the standard form obtained in the Markovian regime, given by 
\ba
\mathcal{I}_{j}(\omega,t) = \pi G_j\left(\omega\right) > 0~.
\label{markI}
\ea

{\bf A minimal quantum thermal machine beyond Markovianity.}~Here we consider as the QHM a two-level system (TLS) WF with states $\ket{0}$ and $\ket{1}$, 
interacting with a hot and a cold thermal bath, described by the Hamiltonian
\ba
\hat{H}(t) = \hat{H}_{\rm S}(t) + \hat{\sigma}_x \otimes \left(\hat{B}_{\rm c} + \hat{B}_{\rm h} \right) + \hat{H}_{\rm B}.
\label{hamiltls}
\ea
The Pauli matrices $\hat{\sigma}_j$ ($j = x, y, z$) act on the TLS,  the operator $\hat{B}_{\rm c}$ ($\hat{B}_{\rm h}$) acts 
on the cold (hot) bath, and $\hat{H}_{\rm B}$ denotes the bath Hamiltonian. 
The resonance frequency $\omega(t)$ of the TLS is sinusoidally modulated by the 
periodic-control Hamiltonian 
\ba
\hat{H}_{\rm S}(t) &=& \frac{1}{2}\omega(t) \hat{\sigma}_z; ~~~~ \sigma_z\ket{1} = \ket{1},~\sigma_z\ket{0} = -\ket{0} \non\\
\omega(t) &=& \omega_0 + \lambda\Delta_{\rm S}\sin\left(\Delta_{\rm S} t \right),
\label{modtls}
\ea
where the relative modulation amplitude is small: $0 < \lambda \ll 1$. The periodic modulation Eq. \eqref{modtls} gives rise
to Floquet sidebands (denoted by the index 
$q = 0, \pm 1, \pm 2, \ldots$) with frequencies 
$\omega_q = \left(\omega_0 + q\Delta_{\rm S}\right)$ and weights $P_q$, which diminish rapidly with increasing 
$|q|$ for  small $\lambda$ (see Methods ``Non-Markovian dynamics of a driven
two-level system in a dissipative bath'') \cite{kofman04unified, kosloff13quantum, klimovsky13minimal}. 

\begin{figure}[H]
\begin{center}
\includegraphics[width= 0.74\columnwidth, angle = 0]{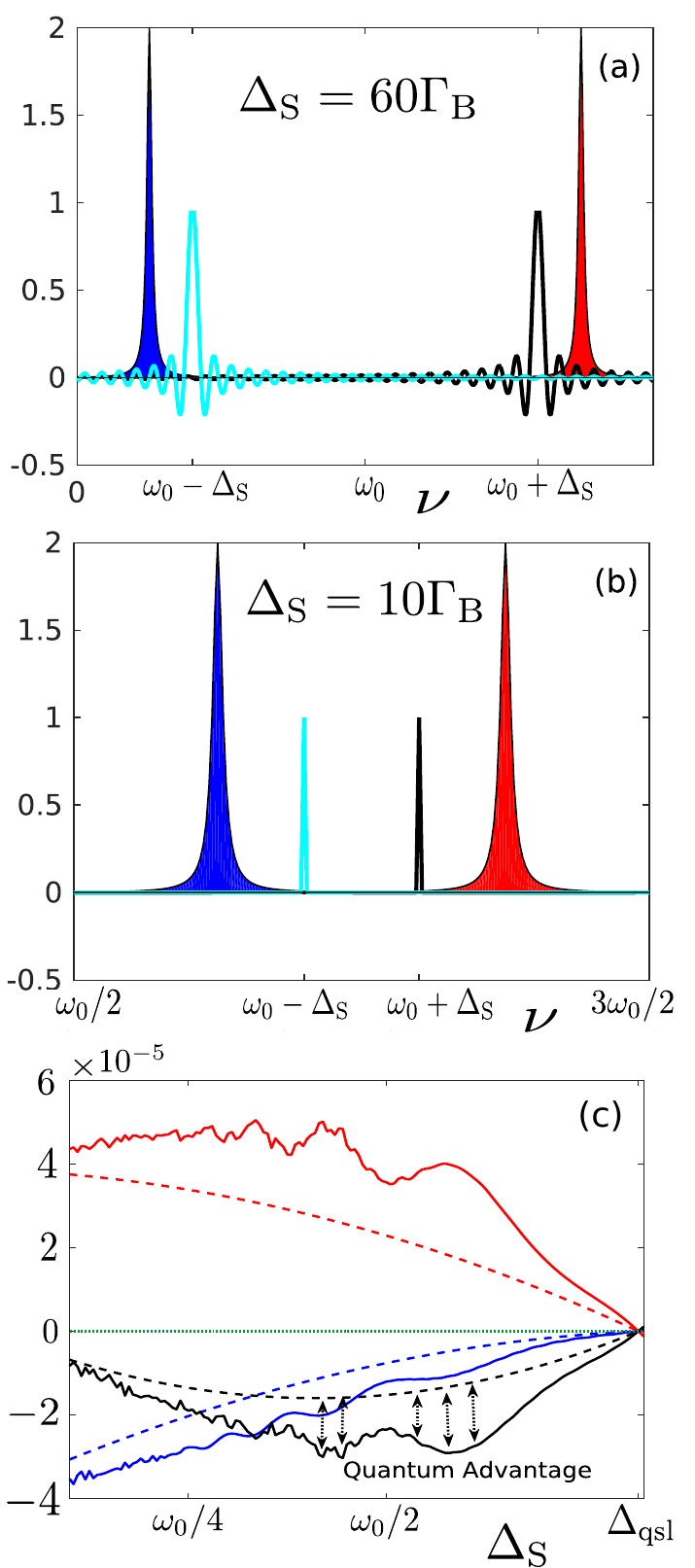}
\end{center}
\caption {{\bf Quantum advantage with quasi-Lorentzian spectral functions:} The quasi-Lorentzian spectral functions of the hot bath $G_{\rm h}(\nu)$ (red filled curve) and the cold bath
$G_{\rm c}(\nu)$ (blue filled curve) (see Eq. \eqref{ghgceq}), and the sinc functions
${\rm sinc} \left[\left(\nu - \omega_0 - \Delta_{\rm S} \right)t\right]$ (black solid curve) and ${\rm sinc} \left[\left(\nu - \omega_0 + \Delta_{\rm S} \right)t\right]$ 
(cyan solid curve) for (a) fast modulation
$\Delta_{\rm S} = 60\Gamma_{\rm B}$ and (b) slow modulation $\Delta_{\rm S} = 10\Gamma_{\rm B}$, at $t = 10\tau_{\rm S}$. Fast (slow) modulation results in broadening (narrowing) of the sinc functions, 
thus leading to enhanced (reduced) overlap with the spectral functions. (c) Power $\overline{\dot{W}}$ (black lines) and heat 
currents $\overline{J_{\rm h}}$ (red lines) and  $\overline{J_{\rm c}}$ (blue lines)
averaged over $n = 10$ modulation periods (solid lines) and the same obtained under the Markovian approximation for 
long cycles, i.e., large
number of modulation periods ($n \to \infty$) (dashed lines), versus the modulation frequency $\Delta_{\rm S}$. Anti-Zeno dynamics  for
$\tau_{\rm C} \lesssim \tau_{\rm B}$ results in output power boost (shown by dotted double-arrowed lines) by up to more than a factor of $2$, signifying quantum advantage  in the heat-engine regime. 
The green dotted line corresponds to zero power and currents.  
Here $\lambda = 0.2, \omega_0 = 20, \gamma_0 = 1, \Gamma_{\rm B} = 0.2, N = 1, \delta = 3, \epsilon = 0.01, \alpha = 1, \beta_{\rm h} = 0.0005, \beta_{\rm c} = 0.005$.}
\label{Gsinc}
\end{figure}
\begin{figure}[H]
\begin{center}
\includegraphics[width= 0.75\columnwidth, angle = 0]{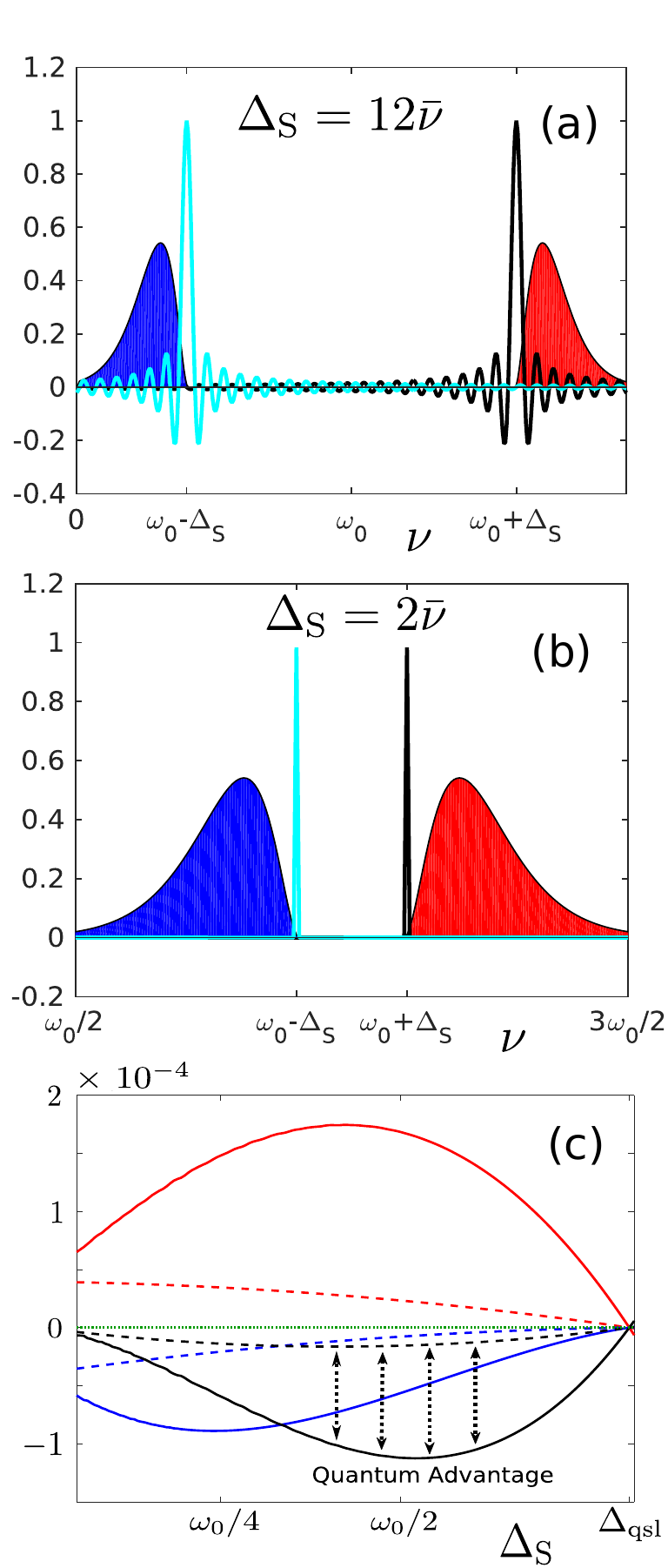}
\end{center}
\caption {{\bf Quantum advantage with super-Ohmic spectral functions:} Overlap of super-Ohmic spectral functions $G_{\rm h}(\nu)$ (red filled curve) and $G_{\rm c}(\nu)$ (blue filled curve) with cutoff
frequency $\bar{\nu}$ (see Eq. \eqref{specso}), with the modulation response 
functions
$ {\rm sinc} \left[\left(\nu - \omega_0 - \Delta \right)t\right]$ (black solid curve) and ${\rm sinc} \left[\left(\nu - \omega_0 + \Delta \right)t\right]$ 
(cyan solid curve) for (a) fast modulation,
$\Delta_{\rm S} = 12 \bar{\nu}$, and (b) slow modulation $\Delta_{\rm S} = 2\bar{\nu}$ at $t = 10\tau_{\rm S}$. Fast (slow) modulation results in broad (narrow) sinc functions, 
and thus enhanced (reduced) overlap with the spectral functions. 
(c) Power $\overline{\dot{W}}$ (black lines) and heat currents $\overline{J_{\rm h}}$ (red lines) and  $\overline{J_{\rm c}}$ (blue lines)
averaged over $n = 10$ modulation periods (solid lines) as compared to the counterparts under Markovian 
approximation for long cycles,  i.e.,  $n \to \infty$  (dashed lines), versus the modulation frequency $\Delta_{\rm S}$.
 A significant quantum advantage is obtained for 
$\tau_{\rm C} \lesssim \tau_{\rm B}$,  when broadening of the sinc functions yields an output power boost (shown by dotted double-arrowed lines) of up to a factor greater than $7$, in the heat engine regime.
The green dotted line corresponds to zero power and currents.  Here $s = 2, \bar{\nu} = 1, \delta = 0.1, \epsilon = 0.1, \alpha = 1, \omega_0 = 20, \gamma_0 = 1, \beta_{\rm h} = 0.0005, \beta_{\rm c} = 0.005$.}
\label{supohm}
\end{figure}

A crucial condition of our treatment is the choice of spectral separation 
of the hot and cold baths, such that the positive sidebands ($q > 0$) 
only couple to the hot bath and the negative 
sidebands 
(with $q < 0$) sidebands  only couple to the cold bath. 
This requirement is satisfied, for example, by the following bath spectral functions:
\ba
G_{\rm h}\left(\omega \right) &=& 0~~\text{for}~ 0 < \omega \leq \omega_0\non\\
G_{\rm c}(\omega) &=& 0~~\text{for}~  \omega \geq \omega_0,
\label{specsep}
\ea
which ensures that for small $\lambda$, only the $q = 1$  harmonic exchanges energy with the hot bath at frequencies 
$\pm\omega_1 = \pm\left(\omega_0 + \Delta_{\rm S}\right)$, while the $q = -1$ harmonic does the same with 
the cold bath at frequencies $\pm\omega_{-1} = \pm\left(\omega_0 - \Delta_{\rm S}\right)$. We neglect the contribution of the higher 
order sidebands ($|q| > 1$) for $0 < \lambda \ll 1$, for which $P_q \to 0$  \cite{kofman04unified, klimovsky13minimal, alicki14quantum, kosloff13quantum, klimovsky15}.
Further, 
we impose the Kubo-Martin-Schwinger (KMS) detailed-balance
condition
\ba
G_j(-\omega) = G_j(\omega)\exp\left(-\omega \beta_j \right),
\label{kms}
\ea
where $\beta_j = 1/T_j$.

For simplicity, 
in what follows,  $G_{\rm h}(\omega)$ and $G_{\rm c}(\omega)$ are assumed to be mutually symmetric around $\omega_0$,
i.e. they satisfy 
\ba
G_{\rm h}(\omega_0 + \nu) = \alpha G_{\rm c}(\omega_0 - \nu)
\label{symm}
\ea
where $\alpha$ is a real positive number and $0 \leq \nu < \omega_0$ (see Methods ``Steady states in the anti-Zeno
dynamics (AZD) regime'').

The WF is first coupled to the thermal baths at an initial time $-t_{\rm{in}}$ ($t_{\rm{in}} \gg \tau_{\rm th} > 0$). 
Irrespective of the value of $\tau_{\rm S}$, at large times $t + t_{\rm in} \gg \tau_{\rm th}$, 
and under the condition of weak WF-baths coupling, one can arrive at a time-independent  non-equilibrium steady state
$\rho_{\rm S} \to \rho_{\rm ss}$ in the energy-diagonal form (see Methods ``Steady states in the anti-Zeno
dynamics (AZD) regime''):
\ba
\rho_{\rm ss} &=& p_{1, {\rm ss}}\ket{1}\bra{1} + p_{0, {\rm ss}}\ket{0}\bra{0}\non\\
\frac{p_{1, {\rm ss}}}{p_{0, {\rm ss}}} &=:& w = \frac{\alpha e^{-\beta_{\rm h}\left(\omega_0 + \Delta_{\rm S}\right) } +  e^{- \beta_{\rm c}\left(\omega_0 - \Delta_{\rm S}\right)}}{1 + \alpha}.
\label{wstate}
\ea
One can then decouple the WF and the baths, such that they are non-interacting for a time interval exceeding $\tau_{\rm B}$ so as 
to eliminate all memory effects, 
then recouple them again at $t = 0$, keeping $\rho_{\rm S} = \rho_{\rm ss}$, and run the QHM in a cycle (as described in the Section ``From Markovian to non-Markovian dynamics'').  

In general, owing to the finite widths ($\sim 1/\tau_{\rm C}$) of $\mathcal{I}_{\rm h, c}(\omega_q, t)$ in the frequency domain for short coupling times ($\tau_{\rm C} \lesssim \tau_{\rm B}$), 
the WF would be driven away from  $\rho_{\rm ss}$, as follows from  Eq. \eqref{rwaTsinc}, causing  $\rho_{\rm S}(t)$ to evolve with time within the time interval $0 < t \leq \tau_{\rm C}$. However,
in order to generate a cyclic QHM operating 
in the steady-state, we focus on cycles consisting of $n$ modulation periods that satisfy  
\ba
\tau_{\rm C}^{-1} \ll T_{{\rm c, h}}; ~~~ \tau_{\rm C}^{-1} < \omega_0 - \Delta_{\rm S},
\label{Tctaus}
\ea
so that 
\ba
e^{-\frac{\omega_0 \pm \Delta_{\rm S} + 1/\tau_{\rm C}}{T_{{\rm c, h}}}} \approx e^{-\frac{\omega_0 \pm \Delta_{\rm S}}{T_{{\rm c, h}}}}.
\label{condtauc}
\ea
The above conditions Eq. \eqref{Tctaus} and \eqref{condtauc}, along with the KMS condition Eq. \eqref{kms}, imply that
\ba
\mathcal{I}_{\rm h}\left(-(\omega_0 + \Delta_{\rm S}), t \right) &\approx& e^{-\frac{\omega_0 + \Delta_{\rm S}}{T_{\rm h}}}\mathcal{I}_{\rm h}\left(\omega_0 + \Delta_{\rm S}, t \right)\non\\
\mathcal{I}_{\rm c}\left(-(\omega_0 - \Delta_{\rm S}), t \right) &\approx& e^{-\frac{\omega_0 - \Delta_{\rm S}}{T_{\rm c}}}\mathcal{I}_{\rm c}\left(\omega_0 - \Delta_{\rm S}, t \right).
\label{kmsI}
\ea
Equation \eqref{kmsI}, in turn, guarantees that Eq. \eqref{wstate} yields the steady state even at short times, and thus eliminates any time dependence in 
$\rho_{\rm S}$ (see Fig. \ref{dyn}). For a QHM operating in the steady 
state,
\ba
\dot{\rho}_{\rm S}(t) = \left( \mathcal{L}_{\rm h} + \mathcal{L}_{\rm c} \right)[\rho_{\rm ss}] \non
\ea
remains zero even during de-coupling from, and re-coupling with the hot and cold baths. This
ensures that the system remains in its steady state $\rho_{\rm ss}$ throughout the cycle.

Equations \eqref{Tctaus} - \eqref{kmsI} can be easily satisfied for  experimentally achievable parameters; 
eg.,  $\Delta_{\rm S} \sim$ kHz, and $n = 10$ would imply $T_{\rm c} \gg \hbar \Delta_{\rm S}/2\pi n k_{\rm B} \sim 10^{-9}$ K. The number $n=10$ was chosen to be around the minimal 
number $n$ that allows for the validity of the secular approximation $q^{\prime} = q$ in \eqref{rwaa} and hence for a simplification \eqref{rwaqq2} in the master equation. This number should be made as 
low as possible, since by decreasing $n$ we decrease  the cycle duration $\tau_{\rm C}$ and hence increase
the power boost, as explained above. Since this power boost is then maximized  without changing the efficiency, as noted above, the performance is optimized for the chosen n.

From
 the First Law of thermodynamics, the QHM output power $\dot{W}(t)$ is given in terms of the hot and cold heat currents $J_{\rm h}(t)$ and 
 $J_{\rm c}(t)$, respectively, by \ct{klimovsky15} 
\ba
\dot{W}(t) = -(J_{\rm h}(t) + J_{\rm c}(t)).
\ea
The possible operational regimes of the heat machine, i.e., its being a heat-engine or a refrigerator \cite{klimovsky13minimal, klimovsky15}, are determined by the signs of the WF-baths coupling duration-averaged
$\overline{J_{\rm h}}$, $\overline{J_{\rm c}}$ and $\overline{W}$. 
One can calculate the steady-state efficiency $\eta$, average power output $\overline{\dot{W}}$ and average heat currents 
$\overline{J_j}$ ($j = {\rm h, \rm c}$)
\ba
\eta &=& -\frac{\oint_{\tau_{\rm C}} \dot{W}(t) dt}{\oint_{\tau_{\rm C}} J_{\rm h}(t) dt}; \non\\
\overline{\dot{W}} &=& \frac{1}{\tau_{\rm C}} \oint_{\tau_{\rm C}} \dot{W}(t) dt;~~~\overline{J_{j}} = \frac{1}{\tau_{\rm C}} \oint_{\tau_{\rm C}} J_j(t) dt
\label{effpw}
\ea
as a function of the modulation speed $\Delta_{\rm S}$, searching for the extrema of the functions in Eq. (\ref{effpw}).

The heat currents  $J_{\rm c}$ and $J_{\rm h}$, flowing out of the cold and hot baths, respectively, are obtained 
consistently with the Second Law \ct{klimovsky13minimal, klimovsky15} in the form 
\ba
J_{\rm h}(t) &=& \frac{\lambda^2}{4}(\omega_{0} + \Delta_{\rm S}) \mathcal{I}_{\rm h}\left(\omega_0 + \Delta_{\rm S},t\right)  \frac{e^{-(\omega_0 + \Delta_{\rm S})\beta_{\rm h}} - w}{w + 1}, \non\\
J_{\rm c}(t) &=& \frac{\lambda^2}{4}(\omega_{0} - \Delta_{\rm S}) \mathcal{I}_{\rm c}\left(\omega_0 - \Delta_{\rm S},t\right)  \frac{e^{-(\omega_0 - \Delta_{\rm S})\beta_{\rm c}} - w}{w + 1},\non\\
~
\label{Jhc}
\ea
where we have used $P_{\pm 1} = \lambda^2/4$.

In order to study the steady-state QHM performances for different modulation frequencies, we consider the example of two non-overlapping spectral 
response functions of the two baths 
 displaced by $\delta$ with respect to $\omega_q$, i.e., $G_{\rm h}(\nu)$ ($G_{\rm c}(\nu)$) characterized by a quasi-Lorentzian peak of width
 $\Gamma_{\rm B}$, with the
 peak at $\nu_{\rm h} = \omega_0 + \Delta_{\rm S} + \delta$ ($\nu_{\rm c} = \omega_0 - \Delta_{\rm S} - \delta$) (see Methods ``Quasi-Lorentzian bath spectral functions''). Alternatively, we also consider the example of 
 two non-overlapping super-Ohmic spectral 
response functions $G_{\rm h}(\nu)$ and $G_{\rm c}(\nu)$ of the two baths, with their origins shifted from $\nu = 0$ by 
$\nu_{\rm h} = \omega_0 + \Delta_{\rm S} - \delta$ and 
$\nu_{\rm c} = \omega_0 - \Delta_{\rm S} + \delta$ respectively, for $0 < \delta \ll \Delta_{\rm S}, \omega_0, \omega_0 - \Delta_{\rm S}$
(see Methods ``Super-Ohmic bath spectral functions'').
The dependence of $\nu_{\rm h, c}$ on $\Delta_{\rm S}$ amounts to considering baths with different spectral functions for different modulation frequencies, and ensures that any enhancement in heat currents and power under fast driving results from
 the broadening (rather than the shift) of the sinc functions, which are centered at $\omega_0 \pm \Delta_{\rm S}$.

\begin{figure}[t]
\begin{center}
\includegraphics[width= 0.85\columnwidth, angle = 0]{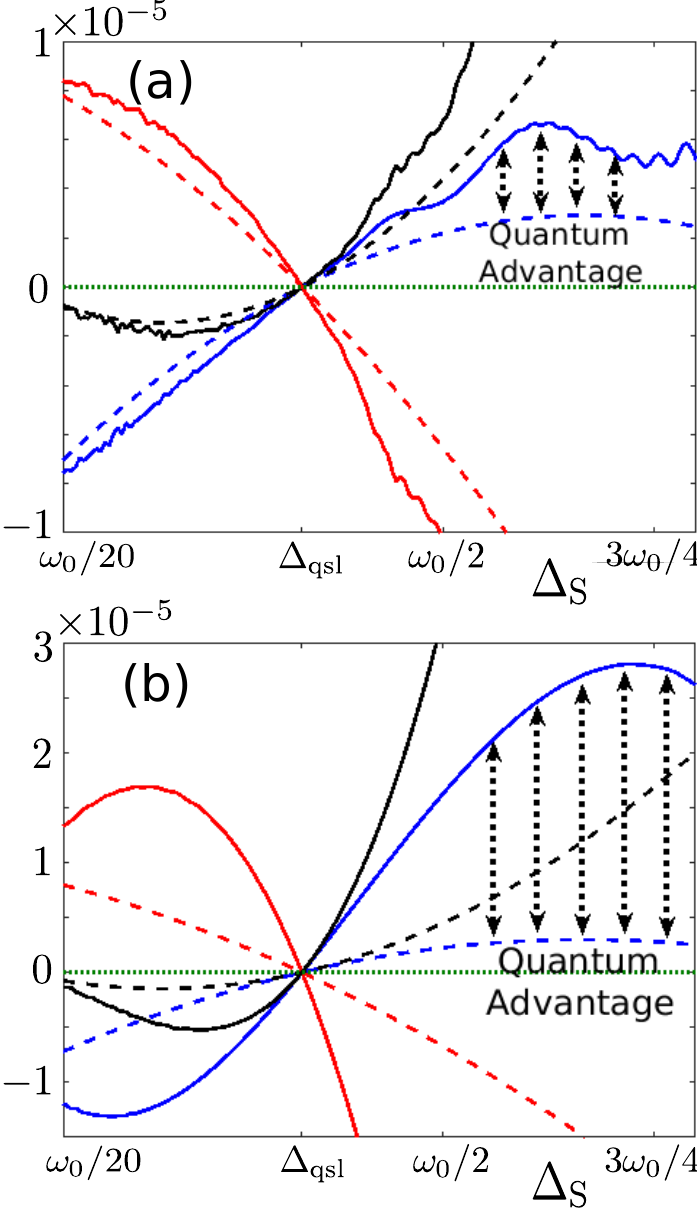}
\end{center}
\caption {{\bf Quantum enhanced refrigeration:} Power $\overline{\dot{W}}$ (black lines) and heat currents $\overline{J_{\rm h}}$ (red lines) and  
$\overline{J_{\rm c}}$ (blue lines)
averaged over $n = 10$ modulation periods (solid lines) as compared to the counterparts under Markovian
approximation for long cycles,  i.e.,  $n \to \infty$  (dashed lines), versus the modulation frequency $\Delta_{\rm S}$, 
 for (a) quasi-Lorentzian spectral functions with $N = 1, \delta = 3, \epsilon = 0.01, \alpha = 1,  \Gamma_{\rm B} = 0.2$  (see Eq. \eqref{ghgceq}) and (b) super-Ohmic spectral functions with 
 $s = 2, \bar{\nu} = 1, \delta = 0.1, \epsilon = 0.1, \alpha = 1$
(see Eq. \eqref{specso}). The enhanced overlap resulting from 
fast modulation (large $\Delta_{\rm S}$) enhances the heat currents $\overline{J_{\rm c}}$ to up to a factor larger than $2$ for (a) and larger than $9$ for (b) in the refrigerator 
regime (shown by dotted double-arrowed lines),
signifying quantum advantage.
The green dotted line corresponds to zero power and currents.  
Here $\lambda = 0.2, \omega_0 = 20, \gamma_0 = 1, \beta_{\rm h} = 0.001, \beta_{\rm c} = 0.002$.}
\label{refr}
\end{figure}

\begin{figure}[t]
\begin{center}
\includegraphics[width= 0.9\columnwidth, angle = 0]{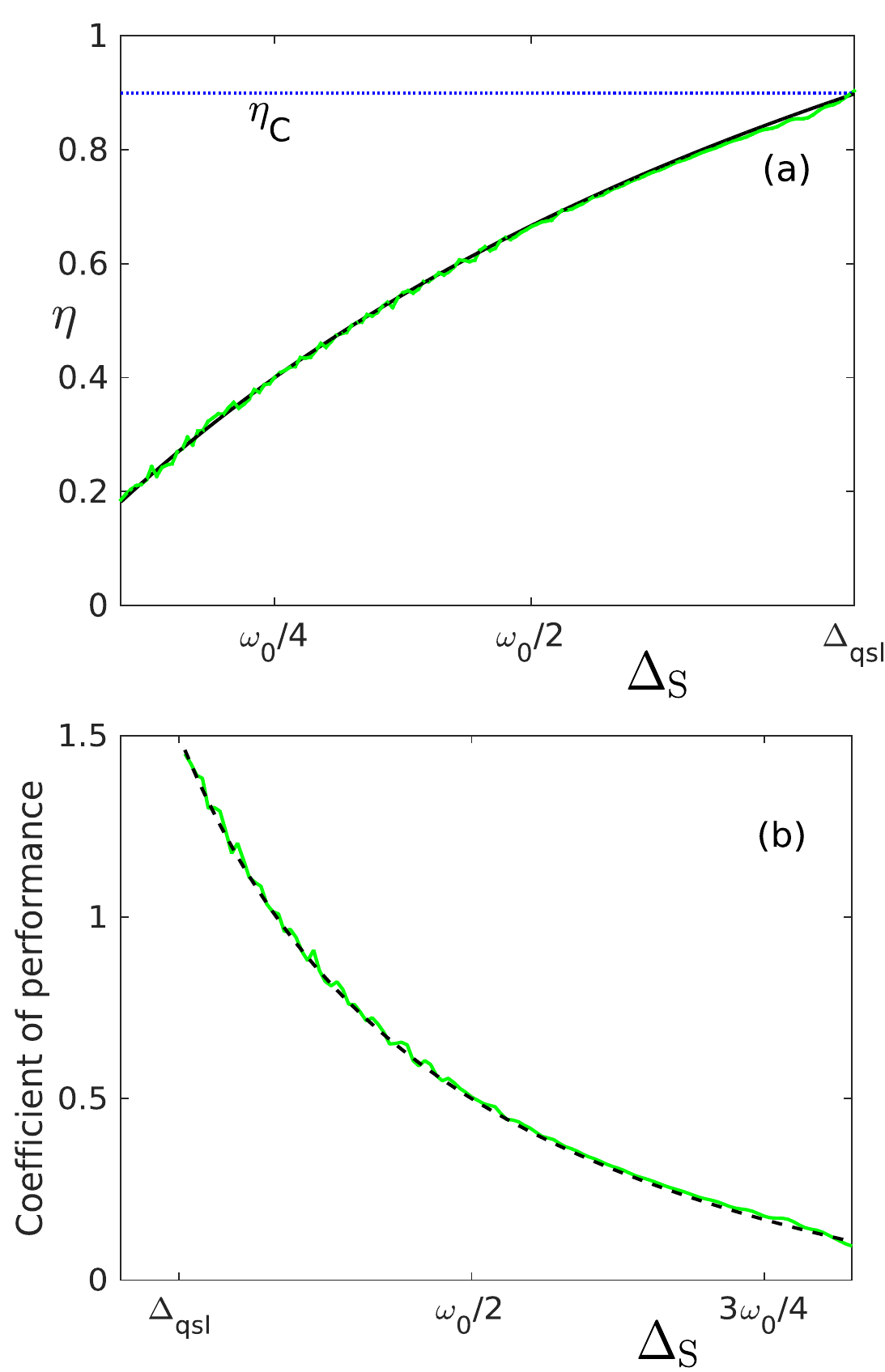}
\end{center}
\caption {{\bf Efficiency and coefficient of performance} (a) Efficiency $\eta$ for the heat engine in the anti-Zeno dynamics regime (green solid line), and the same in
the long coupling time limit $\tau_{\rm C} \to \infty$  (black dashed line), versus the modulation frequency $\Delta_{\rm S}$, for $\beta_{\rm h} = 0.0005$ and $\beta_{\rm c} = 0.005$. 
The efficiency approaches the Carnot limit (blue dotted line) $\eta_{C} = 1 - \beta_{\rm h}/\beta_{\rm c}$ at $\Delta = \Delta_{\rm qsl}$.
(b) Coefficient of performance for the refrigerator in the anti-Zeno dynamics regime (green solid line), and the same in the long time $t \to \infty$ limit (black dashed line), for
$\beta_{\rm h} = 0.001$ and $\beta_{\rm c} = 0.002$. Here $\lambda = 0.2, \omega_0 = 20$, and we consider
thermal machines coupled to thermal baths with quasi-Lorentzian spectral 
functions  with $N = 1, \gamma_0 = 1, \delta = 3, \epsilon = 0.01, \alpha = 1,  \Gamma_{\rm B} = 0.2$ (see Eq. \eqref{ghgceq}).}
\label{effcop}
\end{figure}

We plot quasi-Lorentzian bath spectral functions and the sinc functions in Figs. \ref{Gsinc}a and \ref{Gsinc}b, and the corresponding time averaged heat currents and 
power (see Eq. \eqref{Jhc}) for the heat engine regime in Fig. \ref{Gsinc}c. We do the same for super-Ohmic bath spectral functions in Figs. \ref{supohm}a, \ref{supohm}b and \ref{supohm}c. The corresponding heat currents and powers for the refrigerator regimes are shown in Figs. \ref{refr}a and 
\ref{refr}b.  The  Markovian 
 approximation: ${\rm sinc}(x) \propto \delta(x)$ in 
Eq. \eqref{rwaTsinc} reproduces 
the correct heat currents and power only in the limit of slow modulation ($\tau_{\rm C} \gg \tau_{\rm B}$). By contrast, the Markovian 
 approximation
reproduces the exact efficiency for both slow and fast modulation rates (see Fig. \ref{effcop}a).
Thus, although the efficiency grows as $\tau_{\rm S}$ decreases, it is still limited by the Carnot bound.

{\bf Anti-Zeno dynamics.}~The performance of the QHM depends crucially on the relative width of the spectral function and the sinc functions. A slow modulation 
($\tau_{\rm S} \gg \tau_{\rm B}$) results in sinc 
functions which are non-zero only over a narrow frequency range, wherein
$G_{j}(\nu)$ can be  assumed to be approximately constant, which
  leads to time-independent $\mathcal{I}_j(\omega_q)$ and Markovian dynamics.
On the other hand, fast modulation ($\tau_{\rm C} \lesssim \tau_{\rm B}$) is 
associated with broad sinc
functions, for which $G_{j}(\nu)$ is variable over the frequency range $\sim~\tau_{\rm C}^{-1}$ for which the sinc functions are non-zero  (see Figs. \ref{Gsinc}a, \ref{Gsinc}b, \ref{supohm}a and 
\ref{supohm}b). This regime is 
a consequence of the time-energy 
uncertainty relation of quantum mechanics, and is associated with the anti-Zeno effect \cite{erez08, kofman00acceleration}. This effect results in 
dynamically enhanced system-bath energy exchange, which we dub
anti-Zeno dynamics (AZD). Remarkably, 
in a QHM, appropriate choices of $\mathcal{I}_{\rm h, c}(\omega_q, t)$ may yield a power and heat-currents boost whenever the sinc functions have sufficient overlap with $G_{\rm h, c}(\nu)$ (see Figs. \ref{Gsinc}c 
and  \ref{supohm}c).

Importantly, we find that spectral functions peaked at frequencies sufficiently detuned from $\omega_0 \pm \Delta_{\rm S}$ (i.e., $\delta > \Gamma_{\rm B}$) 
may increase   
 the overlap with the sinc functions appreciably under fast 
 modulation in the anti-Zeno regime, for $\tau_{\rm C}^{-1}, \delta \gtrsim \Gamma_{\rm B}$, thus resulting in substantial
output power boost. This regime indicates that finite spectral width of the sinc functions  may endow a HE with significant quantum advantage,  arising from the time-energy 
uncertainty relation, which is absent in the classical regime,
be it Markovian or non-Markovian.  In the numerical examples shown here, the quantum advantage in the HEs 
powered by baths with quasi-Lorentzian  (super-Ohmic) spectral functions can increase the power by a factor larger than two (seven) (see Figs. \ref{Gsinc}c and \ref{supohm}c), for the same efficiency
(see Fig. \ref{effcop}a and Methods ``Efficiency and coefficient of performance").

{\bf Quantum refrigeration:} AZD can lead to quantum advantage in the refrigerator regime as well,  for modulation rates beyond the quantum speed limit  \cite{mukherjee16speed, deffner17quantum} (see 
App. \ref{app1}),
by enhancing the heat current $\overline{J}_{\rm c}$, thus resulting in 
faster cooling of the cold bath. As for HE, numerical analysis shows that quasi-Lorentzian, as well as  super-Ohmic 
bath spectral functions
can lead to significant quantum advantage
in the AZD regime (see Fig. \ref{refr}).
On the other hand, as for the efficiency in case of the HE, the coefficient of performance
\ba
{\rm COP} = - \frac{\overline{J_{\rm c}}}{\overline{\dot{W}}}
\ea
is not significantly affected by the broadening of the sinc function, 
and on average remains identical to that obtained under slow modulation in the Markovian regime (see Fig. \ref{effcop}b and Methods ``Efficiency and coefficient of performance").

\section*{Discussion} 

We have explored the hitherto uncharted domain of quantum heat engines (QHEs) and refrigerator (QRs) based on quantum working fluids (WFs) intermittently coupled and decoupled from heat baths operating on non-Markovian time scales. We have shown that for driving 
(control) faster than the correlation (memory) time of the bath, one may achieve dramatic output power boost in the anti-Zeno dynamics (AZD) regime.

Let us revisit
our findings,
using as a benchmark the Markovian regime  under periodic driving: In the latter regime, detailed balance of transition rates between the WF levels, as well as the periodic driving  
(modulation) rate, determine, according to the First and Second Laws of thermodynamics, the heat currents between the (hot and cold) baths, and thereby the power produced or consumed. In our 
present treatment, the Markovian regime is recovered under slow modulation, such that the WF-baths coupling duration $\tau_{\rm C}$ exceeds the bath correlation time $\tau_{\rm B}$. 
Then, the Markovian approximation is adequate for studying the operation of the QHE or the QR. By contrast, 
under fast modulations, such that $\tau_{\rm C} = n\tau_{\rm S} \lesssim \tau_{\rm B}$, 
the working fluid interacts with the baths over a broad frequency range of the order of $\sim \tau_{\rm C}^{-1}$, according to the time-energy
uncertainty relation in quantum mechanics. The  frequency-width over which system-bath energy exchange takes place can lead 
to anti-Zeno dynamics (AZD). The resultant quantum advantage is then especially pronounced for bath spectral functions that are appreciably shifted by 
$\delta > \Gamma_{\rm B} \sim \tau_{\rm B}^{-1}$, 
from the centers of the sinc functions that govern the system-bath energy exchange rates.

We have explicitly restricted the results to mutually symmetric bath spectral functions (e.g., the experimentally common Lorentzian or
Gaussian spectra), in order to ensure time-independent steady-states of the WF. Yet this requirement is not essential, since the WF steady-state may be time-dependent
as long as it is periodic so as to allow for cyclic operation. The AZD \cite{kofman01universal, kofman04unified, kofman00acceleration, erez08, gordon09cooling, gordon10equilibration, alvarez10zeno, gordon07universal} can arise for any bath spectra of finite width $\sim 1/\tau_{\rm B}$, as long as $n\tau_{\rm S} \lesssim \tau_{\rm B}$. 
One can therefore operate a thermal machine provided stroke 1 of the cycle is in the AZD regime and achieve a quantum advantage without additional restrictions on the bath spectral functions
(see Methods ``Thermal machines with arbitrary (asymmetric) spectral functions").

The QHM discussed here is driven by external modulation. As previously shown both theoretically \cite{kofman00acceleration, kofman01universal, kofman04unified, gordon07universal, erez08, gordon09cooling}
and experimentally \cite{alvarez10zeno, almog11direct},  periodic perturbations of the TLS state  
can increase its relaxation rate  in the  non-Markovian anti-Zeno regime. The reason for the power boost is that at the non-Markovian stage of the evolution which occurs on short time scales,
the sinc factors in the convolutions with $G(\omega)$, as in \eqref{Iazd}, are sufficiently broad so as to modify the convolutions and hence the relaxation rates in \eqref{rateq} in comparison with
the Markovian 
case, where these sinc functions are spectrally narrow enough to be approximated by delta-functions. Under the conditions chosen in the paper, this modification leads to an increase in the TLS 
relaxation rates and hence to a power boost. This boost is of quantum nature, since the broadening of the 
sinc factors is due to the quantum time-energy uncertainty relation that may lead to the violation of energy conservation at short times. The quantum mechanical 
time-energy uncertainty relation employed here reflects the fact that the Scr\"{o}dinger equation for a two-level system coupled to a  bath renders the energy transfer 
probability from the two-level system to the bath and back oscillatory in time. Such oscillation leads at short times (comparable to the required cycle period) to 
sinc-like deviation from delta-function energy conservation. Classical description of analogous processes, even beyond the Markovian approximation, does not involve 
discrete energy levels and hence no oscillations of the system-bath transfer rate that deviates from energy conservation. Thus, the effects discussed here are inherently quantum mechanical.

The non-Markovian effect in the present context
is quantified by the  spectral widths of the sinc functions compared to the bath-response $G(\omega)$  spectral width $1/\tau_{\rm B}$. If the cycle duration is kept fixed, 
then the non-Markovian effect scales with the 
spectral width of $G(\omega)$. Hence, super-Ohmic
bath spectra with their salient cutoff provide realistic examples of the non-Markovian effects described here. Such 
bath spectra should be contrasted with the flatter and broader Ohmic spectra. Yet, non-Markovian dynamics does not necessarily imply a quantum advantage, as discussed in 
App. \ref{app2}.

The predicted power boost relies on transient dynamics: the heat fluxes change with time within $t= \tau_C$ in the non-Markovian 
AZD regime, even when the WF state hardly  changes during that time interval. Yet it is essential that we incorporate
this transient dynamics within steady-state cycles by  decoupling the WF from the baths, allowing the bath-correlations to vanish within $\tau_{\rm B}$ and then recoupling the WF again to the baths when
they have all resumed their initial states. These cycles can be repeated without restriction, thereby allowing us to operate the QHM with quantum enhanced performance even for long times,
despite the reliance on transient dynamics within the stroke 1 of each cycle.

The quantum advantage of AZD, at zero energetic cost (see App. \ref{app3}), manifests 
itself in the form of higher output power, for the same efficiency, in the QHE regime ($\Delta_{\rm S} < \Delta_{\rm qsl}$), as compared to that obtained under Markovian dynamics in the limit of large $\tau_{\rm C}$, 
all other parameters remaining the same. 
Alternatively,  in the 
QR regime ($\Delta_{\rm S} > \Delta_{\rm qsl}$), AZD may lead to quantum advantage over Markovian dynamics in the form of higher heat current $\overline{J}_{\rm c}$, or, equivalently, higher cooling rate 
of the cold bath, for the same 
coefficient of performance.  The latter effect leads to the enticing possibility of quantum-enhanced speed-up of the cooling rate of systems as we approach the absolute zero, and raises
questions regarding the 
validity of the Third Law of Thermodynamics in the quantum non-Markovian regime, if we expect the vanishing of the cooling rate at zero temperature as a manifestation of the Third 
Law \cite{kolar12, paz17, masanes17a}.  

The QHE power boost in the anti-Zeno regime results from a corresponding increase in the rates of heat-exchange and entropy
production, arising from the TLS relaxation by both baths. This is the reason that the efficiency, i.e., the ratio of the work output to the heat input, is
unchanged, i.e. is the same as in the standard Markovian regime. Yet, all parameters being equal, the QHM rate of operation (as measured by the power
output) speeds up in the anti-Zeno regime, which constitutes a practical quantum advantage.

One can extend the analysis discussed here to Otto cycles \cite{mukherjee16speed, kosloff17the}: Fast periodic modulation during the non-unitary strokes of an Otto cycle can 
speed up the thermalization through AZD, thereby allowing quantum enhanced performance.  Interestingly, fast modulation in the Otto cycle can yield
enhanced power or refrigeration rate, even in the Markovian regime \cite{erdman18maximum}.

\begin{figure}[t]
\begin{center}
\includegraphics[width= \columnwidth, angle = 0]{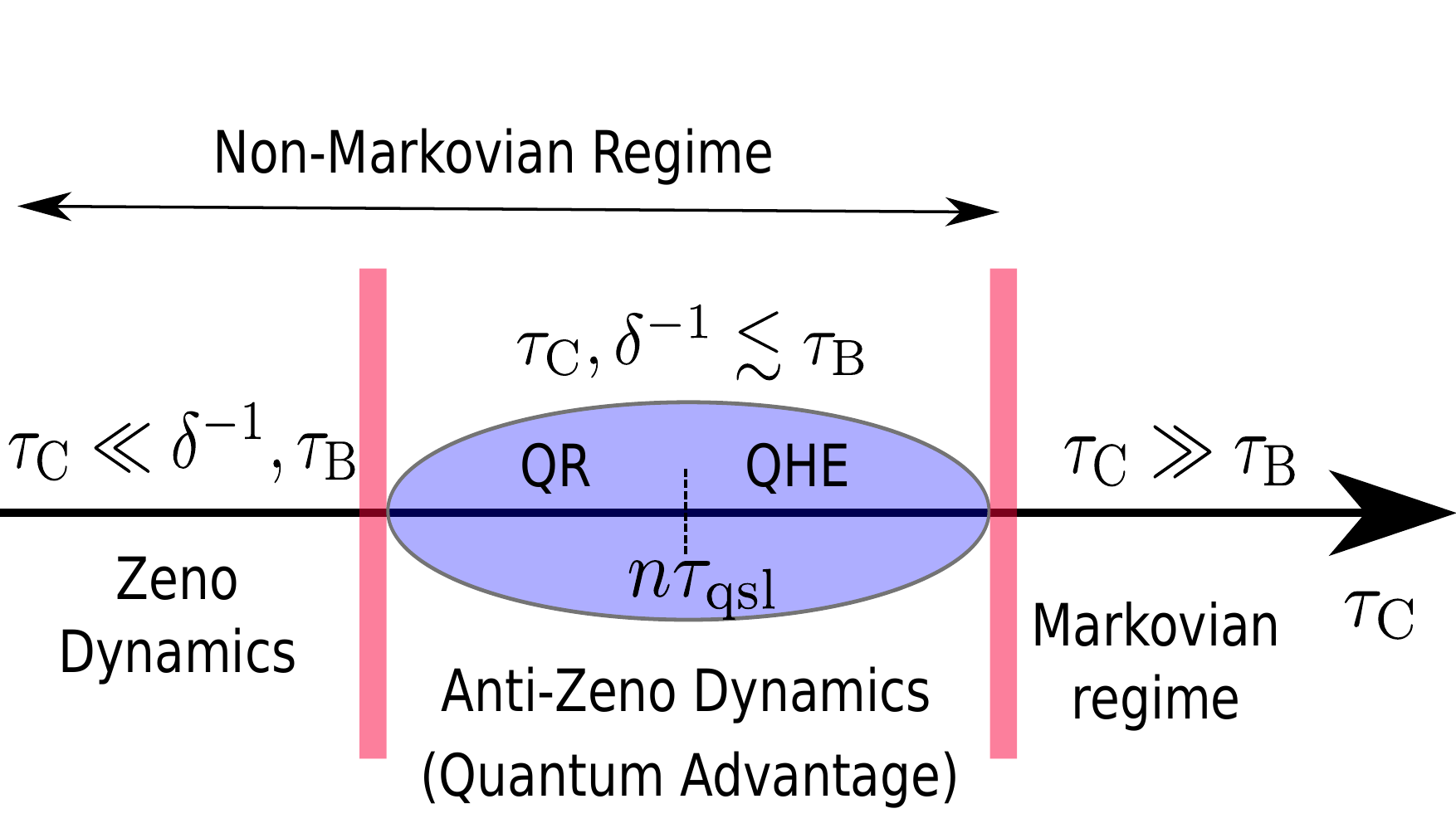}
\end{center}
\caption {{\bf Dynamical regimes:} Schematic display of the different regimes of operation, as a function of the working fluid (system) - baths coupling duration $\tau_{\rm C}$.}
\label{regimes}
\end{figure}
\begin{figure}[h]
\begin{center}
\includegraphics[width= 0.7\columnwidth, angle = 0]{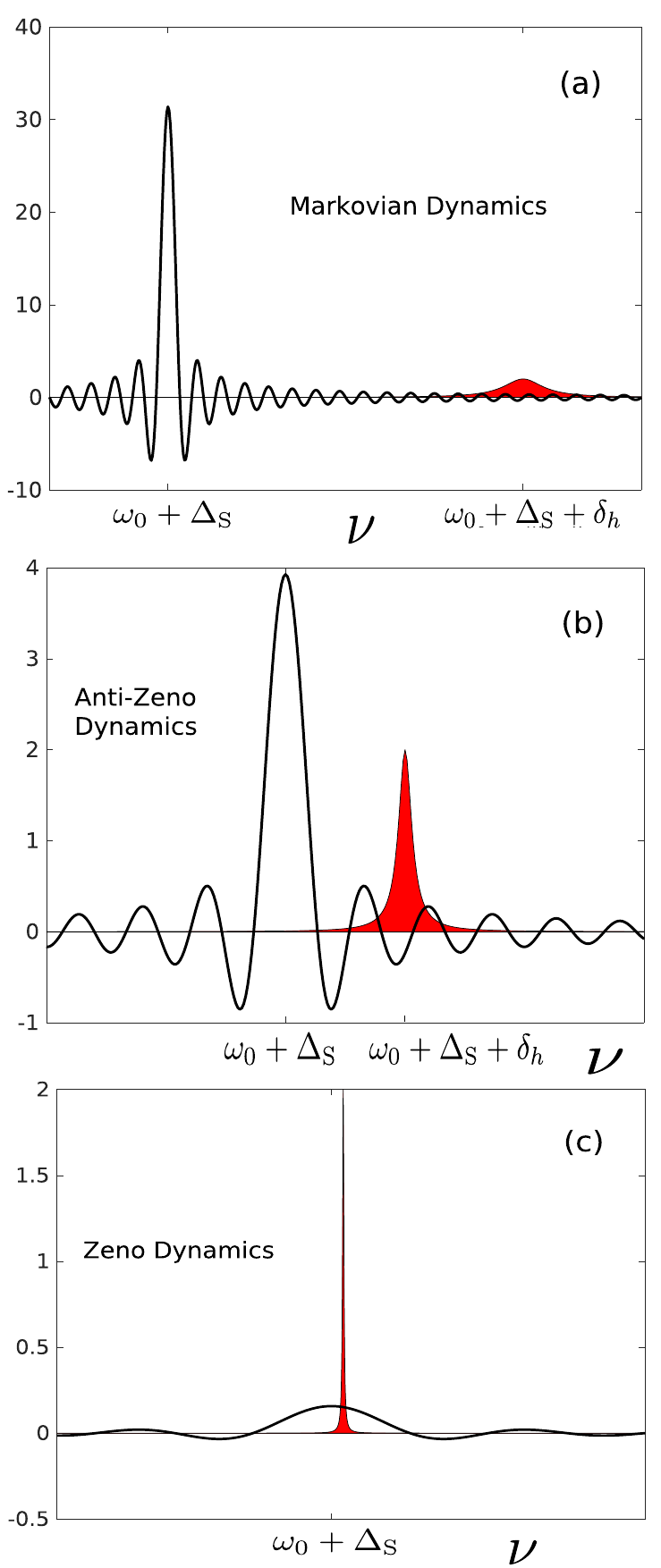}
\end{center}
\caption {{\bf Spectral and sinc functions:} The quasi-Lorentzian spectral function $G_h(\nu)$ (red filled curve)
and ${\rm sinc} \left[\left(\nu - \omega_0 - \Delta \right)t\right]$ (black solid curve) at time $t = 10\tau_{\rm S}$ for (a) Markovian dynamics with 
$\Delta_{\rm S} = 10\Gamma_{\rm B}$, (b) anti-Zeno dynamics with $\Delta_{\rm S} = 80\Gamma_{\rm B}$ and (c) Zeno dynamics with $\Delta_{\rm S} = 2000\Gamma_{\rm B}$.
The overlap between the two functions
is maximal for anti-Zeno dynamics, while it vanishes both for slow modulation, when the sinc function is narrow, as well as for Zeno dynamics, when the sinc function is much broader than $G_h(\nu)$, and approaches zero 
for all $\nu$. The same $G_h(\nu)$ is considered in (a), (b) and (c). Here $\omega_0 = 20, \gamma_0 = 1, \Gamma_{\rm B} = 0.2, N = 1, \delta_{\rm h} = 3, \epsilon = 0.01, \alpha = 1$.}
\label{zazdreg}
\end{figure}

Finally, in the regime of ultrafast modulation with $\tau_{\rm C}^{-1}\gg \Gamma_{\rm B}, \delta$,  quantum Zeno dynamics sets in,
 leading to vanishing heat currents and power, thus
implying that such a regime is incompatible with thermal machine operation (see Fig. \ref{regimes},  App. \ref{app2} and  Fig. \ref{zazdreg}). While Zeno dynamics has
commonly  been associated with
measurements \cite{misra77the, kofman00acceleration, itano90quantum, kofman01zeno}, both the Zeno and the anti-Zeno effects occur under various frequent perturbations, such as phase flips and 
nonselective (unread) measurements. Generally speaking, to observe the discussed effects, it is sufficient to repeatedly perform cycles of coupling the system (here - the WF) with another system
(here - a bath), then destroying or sharply changing the coherence between the two. This sharp change can be effected in different ways, e.g., by a measurement of the system (which can be read-out or not)
or, as in our case, by abruptly decoupling the WF and the baths, which gives rise to Zeno or anti-Zeno 
dynamics \cite{kofman01universal, kofman04unified, kofman00acceleration, erez08, gordon09cooling, gordon10equilibration}.
In contrast to previous studies, here the Zeno or anti-Zeno dynamics of the WF arises by an external periodic field, and therefore not around the frequency $\omega_0$ of the unperturbed system, 
as in previous cases, but at multiple sideband frequencies $\omega_0 + q \Delta_{\rm S}$.

The Markovian approximation suffices to find the correct efficiency (for a QHE) or the
coefficient of performance (for a QR), even in the non-Markovian regimes, for mutually symmetric bath spectral functions (see Eq. \eqref{symm}).
 This guarantees that the efficiency always remains
below the Carnot bound, even under 
fast modulations (see Fig. \ref{effcop}). 

Our scenario is conceptually different from that in which work is produced by a QHE on an external quantum system and
quantum effects arise from the
interaction between the quantum WF and the external quantum system  \cite{watanabe17quantum}. Such quantum effects are absent in our case, where work input in the QHE is provided by a classical field.

Experimental scenarios where the predicted AZD quantum advantage may be tested are diverse. Since non-Markovianity in general, and AZD in particular, 
require non-flat bath spectral functions, suitable candidates for the hot and cold baths are microwave cavities and waveguides in which dielectric gratings are embedded, with distinct cut-off and 
bandgap frequencies \cite{magnusson92new, klimovsky15} and the WF is a qubit whose level distance is modulated by fields. The required qubit modulations are then compatible with MHz periodic 
driving of superconducting transmon qubits \cite{Houck08controlling, Peterer15coherence} 
or NV-center qubits in diamonds \cite{sangtawesin14fast}. One can effectively decouple the WF from the thermal baths in stroke 2 by abruptly 
changing the resonance frequency of the two-level WF from $\omega_0$ to $\tilde{\omega}$, thus rendering the WF strongly off-resonant with the thermal baths, so that
$G_{\rm h}\left(\tilde{\omega}\right) \approx G_{\rm c}\left(\tilde{\omega}\right) = 0$), thereby precluding any energy flow between the baths and the WF. We can recouple the WF with the 
thermal baths by reverting this frequency back to $\omega_0$, and then modulating it periodically, so as to generate either Markovian or non-Markovian anti-Zeno dynamics, 
as discussed above \cite{alvarez10zeno}.

The AZD regime was experimentally observed in NMR setups \cite{alvarez10zeno}. 
Micro/nano-scale heat machines have been experimentally realized, 
for a trapped calcium ion as the WF \cite{rossnage16a}; nano-mechanical oscillators WF powered by squeezed thermal bath \cite{klaers17squeezed}; atomic heat machines assisted by quantum
coherence \cite{klatzow19experimental}; or a nuclear spin $1/2$ as the WF in a quantum Otto cycle  \cite{peterson18experimental}.

The novel effects and performance trends of QHE and QR in the non-Markovian time domain, particularly the anti-Zeno induced power boost, open new, 
dynamically-controlled pathways in the quest for genuine quantum features
in heat machines, which has been a major motivation of quantum thermodynamics in recent 
years \cite{ghosh17catalysis, ghosh2018thermodynamic, scully03extracting, fialko12isolated, berut12experimental, rossnage14nanoscale, uzdin15equivalence, klaers17squeezed, klimovsky18single, niedenzu18quantum, ghosh18we, binder18book}.

\section*{Methods}

{\bf Floquet Analysis of the non-Markovian Master Equation.}~Let us consider the differential non-Markovian master equation for the system density operator $\rho_{\rm S}(t)$ in the interaction
picture \cite{kofman04unified}:
\ba
\dot{\rho}_{\rm S}(t) &=& - \int^t_0 ds {\rm Tr}_{\rm B}\big[\hat{S}(t)\otimes \hat{B}_{\rm c}(t) \non\\ 
&+& \hat{S}(t)\otimes \hat{B}_{\rm h}(t), [\hat{S}(s)\otimes \hat{B}_{\rm c}(s) \non\\
&+& \hat{S}(s)\otimes \hat{B}_{\rm h}(s), \rho_{\rm S}(t)\otimes \rho_{\rm B}]\big].
\label{rhosdot}
\ea 
Here $\rho_{\rm B} = \rho_{\rm Bc}\otimes\rho_{\rm Bh}$, where $\rho_{{\rm B}j}$ is the density operator of bath $j$. 
In the derivation of Eq. \eqref{rhosdot} we have assumed that
Tr$[\hat{B}_j,\rho_{{\rm B}j}]=0$. We consider commuting bath operators 
$\left[\hat{B}_{\rm c}(t), \hat{B}_{\rm h}(t^{\prime})\right] = 0$, such that the two baths act additively in Eq. \eqref{rhosdot}.
Below we focus on only one of the baths and omit the labels $c/h$ for simplicity. We then have
\ba
\hat{S}^{\dagger}(t) &=& \hat{S}(t)\non\\
\hat{B}^{\dagger}(t) &=& \hat{B}(t)\non\\
{\rm Tr}\left[\hat{B}(t) \hat{B}(s) \rho_{\rm B}\right] &=& \langle \hat{B}(t) \hat{B}(s) \rangle \equiv \Phi(t-s)\non\\
\hat{S}(t) &=& \sum_{q,\omega} S_{q,\omega} e^{-i(\omega +  q\Delta_{\rm S})t}.
\label{misc}
\ea
where $q$ are integers and $\omega$ are transition frequencies of the system $\mathcal{S}$.

One can use Eq. (\ref{misc}) to write the first term on the r.h.s. of Eq. (\ref{rhosdot}) as
\ba
T_1 &=& - \sum_{\omega, \omega^{\prime}, q, q^{\prime}}  e^{i[(\omega^{\prime} - \omega) +  (q^{\prime} - q)\Delta_{\rm S}] t}\hat{S}^{\dagger}_{q^{\prime},\omega^{\prime}}\hat{S}_{q,\omega} \rho_{\rm S}(t) \non\\
&\times&\int^{t}_0 \big[ \Phi(t-s)  e^{i(\omega +  q\Delta_{\rm S})(t-s)}\big] ds.
\label{rwaT1}
\ea

In the limit of times of interest, i.e., times larger than the period of driving $\tau_{\rm S}$ and the effective periods of the system, $t\gg\tau_{\rm S},(\omega + q\Delta_{\rm S})^{-1}$,
the terms with the fast oscillating factor before the integral in Eq. \eqref{rwaT1} become small and can be neglected, i.e., the secular approximation 
becomes applicable, such that
\ba
\left(q^{\prime} - q\right)\Delta_{\rm S} &=& \omega - \omega^{\prime},
\label{rwacond}
\ea
which generally holds only for
\ba
\omega^{\prime} &=& \omega;~~q^{\prime} = q,
\label{rwaa}
\ea
as long as $\left(q^{\prime} - q\right)\Delta_{\rm S}$ is not close to $\left(\omega^{\prime} - \omega\right)$ for any $q, q^{\prime}, \omega, \omega^{\prime}$.
Condition (\ref{rwaa}) 
gives us
\ba
T_1 &\approx& -\sum_{\omega, q} \hat{S}^{\dagger}_{q,\omega}\hat{S}_{q,\omega} \rho_{\rm S}(t)\non\\
&&\int^{t}_0 \big[ \Phi(t-s)  e^{i(\omega +  q\Delta_{\rm S})(t-s)}\big] ds \non\\
&=& -\sum_{\omega, q} \hat{S}^{\dagger}_{q,\omega}\hat{S}_{q,\omega} \rho_{\rm S}(t)\int^{t}_0 \big[ \Phi(\mu)  e^{i(\omega +  q\Delta_{\rm S})\mu}\big] d\mu \non\\
&=& -\sum_{\omega, q} \hat{S}^{\dagger}_{q,\omega}\hat{S}_{q,\omega} \rho_{\rm S}(t) \non\\
&\times&\int_{-\infty}^{\infty}  G(\nu)  \int^{t}_0  e^{-i \left[\nu - \left(\omega + q\Delta_{\rm S}\right)\right]\mu} d\mu d\nu ,
\label{rwaqq2}
\ea
where $\mu = t - s$, and  
\ba
\Phi(\mu)=\int_{-\infty}^\infty d\nu G(\nu)e^{-i\nu \mu}.
\label{phiG}
\ea

In the limit of slow modulation, such that $t = n\tau_{\rm S} \gg \tau_{\rm B}$, one can perform the Markov approximation, thereby extending the upper
limit of the integral in time in Eq. \eqref{rwaqq2} to $t \to \infty$,
which finally results 
in the time-independent Markovian form \cite{breuer02}
\ba
T_1 &\approx& -\pi\sum_{\omega, q\geq 0} \hat{S}^{\dagger}_{q,\omega}\hat{S}_{q,\omega} \rho_{\rm S}(t) G(\omega,q).
\label{tim}
\ea
On the other hand, in the limit of $t\sim n\tau_{\rm S} \lesssim \tau_{\rm B}$, the Markovian approximation becomes invalid, and one gets 
\ba
T_1 &\approx& -\sum_{\omega, q} \hat{S}^{\dagger}_{q,\omega}\hat{S}_{q,\omega} \rho_{\rm S}(t)\int_{-\infty}^{\infty} d\nu G(\nu) \non\\
&&\Big[\frac{\sin\left(\left[\nu - \left(\omega_0 + q\Delta_{\rm S} \right) \right]t\right)}{\nu - \left(\omega_0 + q\Delta \right)} \non\\
&+& i \left(\frac{\cos\left(\left[\nu - \left(\omega_0 + q\Delta_{\rm S} \right) \right]t\right) - 1}{\nu - \left(\omega_0 + q\Delta_{\rm S} \right)}\right)\Big].
\label{rwanm}
\ea

Progressing similarly as above, one can arrive at similar expressions for other terms in Eq. \eqref{rhosdot} as well.

{\bf Non-Markovian dynamics of a driven two-level system in a dissipative bath.}~The non-Markovian master equation followed by the TLS WF subjected to the Hamiltonian Eq. \eqref{hamiltls} is (see Eq. \eqref{rhosdot})
\begin{widetext}
\ba
\mathcal{L}\left[{\rho}_s(t)\right] &=& \left[\left(A_{\downarrow} + \bar{A}_{\downarrow}\right)\sigma^{+}\rho_s(t)\sigma^{-} - A_{\downarrow}\sigma^{-}\sigma^{+}\rho_s(t) - \bar{A}_{\downarrow}\rho_s(t)\sigma^{-}\sigma^{+}\right] \non\\
&+& \left[\left(A_{\uparrow} + \bar{A}_{\uparrow}\right)\sigma^{-}\rho_s(t)\sigma^{+} - A_{\uparrow}\sigma^{+}\sigma^{-}\rho_s(t) - \bar{A}_{\uparrow}\rho_s(t)\sigma^{+}\sigma^{-}\right] + M \sigma^{-}\rho_s(t)\sigma^{-} + \bar{M}\sigma^{+}\rho_s(t)\sigma^{+}, 
\label{Ltls}
\ea
where we have removed the $h, c$ indices for simplicity, and considered the dynamics due to a single bath. Here
\ba
A_{\downarrow} &=& \sum_{q,q^{\prime} \in \mathbb{Z}} \xi(q^{\prime})\overline{\xi}(q) e^{i\left(q - q^{\prime} \right)\Delta_{\rm S} t} \int^{\infty}_{-\infty}G(\nu)\int^t_0 e^{-i\left[\nu - \left(\omega_0 + q\Delta_{\rm S} \right) \right]\tau} d\nu d\tau, \non\\
A_{\uparrow} &=& \sum_{q,q^{\prime}\in  \mathbb{Z}}\overline{\xi}(q^{\prime})\xi(q) e^{-i\left(q - q^{\prime} \right)\Delta_{\rm S} t}\int^{\infty}_{-\infty}G(\nu)\int^t_0 e^{-i\left[\nu + \left(\omega_0 + q\Delta_{\rm S} \right) \right]\tau} d\nu d\tau, \non\\
M &=& \sum_{q,q^{\prime}\in  \mathbb{Z}} \xi(q) \xi(q^{\prime}) e^{-i\left[2\omega_0 + \left(q + q^{\prime}\right)\Delta_{\rm S}\right]t} \int^{\infty}_{-\infty}G(\nu)\left(\int^t_0 e^{i\left[\nu - \left(\omega_0 + q\Delta_{\rm S} \right)\right]\tau} d\tau +  \int^t_0 e^{-i\left[\nu + \left(\omega_0 + q\Delta_{\rm S} \right)\right]\tau} d\tau\right)d\nu, \non\\
\sigma_x(t) &=& \sum_{q\in  \mathbb{Z}} \left(\xi(q) e^{-i\left(\omega_0 + q\Delta_{\rm S}\right)t}\sigma^{-} + \overline{\xi}(q) e^{i\left(\omega_0 + q\Delta_{\rm S}\right)t}\sigma^{+}\right),\non\\
\xi(q) &=& \frac{1}{\tau_{\rm S}} \int_0^{\tau_{\rm S}} e^{i\int^{t}_0 \lambda \Delta_{\rm S}\sin\left(\Delta_{\rm S} t^{\prime}\right) dt^{\prime}} e^{i q \Delta_{\rm S} t} dt
\ea
\end{widetext}
$\bar{A}_{\downarrow}$, $\bar{A}_{\uparrow}$ and $\bar{M}$ are the complex conjugates of $A_{\downarrow}$, $A_{\uparrow}$ and $M$, respectively. The terms corresponding to $\sigma^{\pm}\rho_{\rm S}(t)\sigma^{\pm}$ in 
Eq. \eqref{Ltls} vanish for diagonal steady-state $\rho_{\rm S}(t) \to \rho_{\rm ss}$ (see Eq. \eqref{wstate}).
Here we focus on times longer than several modulation periods, i.e., $t = n\tau_{\rm S} \gg \tau_{\rm S}$, when the fast oscillatory terms corresponding to
$q \neq q^{\prime}$ vanish as 
well, such that 
\ba
A_{\downarrow} &\approx& \sum_{q} P_q \int^{\infty}_{-\infty}G(\nu)\int^t_0 e^{-i\left[\nu - \left(\omega_0 + q\Delta_{\rm S} \right) \right]\tau} d\nu d\tau, \non\\
A_{\uparrow} &\approx& \sum_{q} P_q \int^{\infty}_{-\infty}G(\nu)\int^t_0 e^{-i\left[\nu + \left(\omega_0 + q\Delta_{\rm S} \right) \right]\tau} d\nu d\tau, \non\\
P_q &=& |\xi(q)|^2.
\ea

We note that 
\ba
&&\int^t_0 e^{\pm i\left[\nu \pm \left(\omega_0 + q\Delta_{\rm S} \right) \right]\tau}  d\tau = \frac{\sin\left(\left[\nu \pm \left(\omega_0 + q\Delta_{\rm S} \right) \right]t\right)}{\nu \pm \left(\omega_0 + q\Delta_{\rm S} \right)} \non\\
&\mp& i \left[\frac{\cos\left(\left[\nu \pm \left(\omega_0 + q\Delta_{\rm S} \right) \right]t\right) - 1}{\nu \pm \left(\omega_0 + q\Delta_{\rm S} \right)}\right].
\label{reim}
\ea
The imaginary part in Eq. \eqref{reim} acts on terms of the form 
$i {\rm Im}[\tilde{\mathcal{I}}_j\left(\pm \omega_q, t \right)]\left(\sigma^{\mp}\sigma^{\pm}\rho_{\rm S}(t) - \rho_{\rm S}(t)\sigma^{\mp}\sigma^{\pm}\right)$, which vanish at large times when
the off-diagonal elements $\rho_{\rm S}(t)$ approach zero
 for any 
initial state. 
On the other hand, the real part of 
Eq. \eqref{reim} gives rise to terms of the form
\ba
\mathcal{I}_j\left(\pm \omega_q, t \right) &:=& {\rm Re}[\tilde{\mathcal{I}}_j\left(\pm \omega_q, t \right)]  \non\\
&=& \int^{\infty}_{-\infty}  G_j(\nu)   \frac{\sin\left(\left[\nu \mp \left(\omega_0 + q\Delta_{\rm S} \right) \right]t\right)}{\nu \mp \left(\omega_0 + q\Delta_{\rm S} \right)} d\nu,
\label{Iazd}
\ea
In the limit of slow modulation such that $t \sim n\tau_{\rm S} \gg \tau_{\rm B}$ ($n \in \mathbb{Z},~ n \gg 1$), the function
$\sin\left(\left[\nu \pm \left(\omega_0 + q\Delta_{\rm S} \right) \right]t\right)/\left[\nu \pm \left(\omega_0 + q\Delta_{\rm S} \right)\right]$ 
assumes a delta function centred at $\nu = \pm\left(\omega_0 + q\Delta_{\rm S}\right)$, thus leading to 
the familiar Markovian form of 
master equation, with 
\ba
\mathcal{I}_j\left(\pm \omega_q, t \right)  = \pi G_j\left[\pm\left(\omega_0 + q\Delta_{\rm S}  \right)\right]~~\forall~t.
\label{Imkv}
\ea
On the other hand, in the anti-Zeno regime of fast modulation: $t \sim n\tau_{\rm S} \lesssim \tau_{\rm B}$,  
$\mathcal{I}_j\left(\pm \omega_q, t \right)$ is not given by Eq. \eqref{Imkv}, and one needs to consider the full form Eq. \eqref{Iazd}.

In particular, for a diagonal state $\rho_{\rm S}(t) = p_1(t)\ket{1}\bra{1} + p_1(t)\ket{0}\bra{0}$, the dynamics Eq. \eqref{Ltls} - Eq. \eqref{Iazd} leads us to the rate equations
\ba
\dot{p}_1(t) &=& -\dot{p}_0(t) = R_0(t)p_0(t) - R_1(t)p_1(t)\non\\
R_0(t) &=&  \frac{\lambda^2}{4}\left[\mathcal{I}_{\rm h}(-\omega_0-\Delta_{\rm S},t) + \mathcal{I}_{\rm c}(-\omega_0+\Delta_{\rm S},t)\right]\non\\
R_1(t) &=&  \frac{\lambda^2}{4}\left[\mathcal{I}_{\rm h}(\omega_0+\Delta_{\rm S},t) + \mathcal{I}_{\rm c}(\omega_0-\Delta_{\rm S},t)\right]
\label{rateq}
\ea

In the Zeno regime of ultra-fast modulation, obtained in the limit of $t \sim n\tau_{\rm S} \ll \tau_{\rm B}$,
the integral $\mathcal{I}(\omega_q,t)$ vanishes (see Fig. \ref{zazdreg}), thus leading to 
the Zeno effect of no dynamics.

Equations \eqref{rhosdot},  \eqref{rwaqq2} and \eqref{Ltls} are  of the type known as the differential master equation (DME). An alternative approach is based on the (less convenient)  integro-differential 
master equation (IME). The two equations are mathematically different and hence require different procedures for reducing them to the Markovian master equation (MME). However, the IME and the
DME have the same validity conditions, i.e., generally similar accuracy. The IME and the DME follow from the exact expansions in
the totally ordered and partially ordered cumulants, respectively, upon neglecting terms of order higher than $2$ in the system-bath coupling, which determines their
accuracy \cite{breuer02, kofman90non, gordon07universal}

The rates \eqref{rateq} can be negative when the modulation (or measurement) period is short enough to break the rotating wave approximation (RWA). Yet the probabilities ${p}_0(t)$, ${p}_1(t)$ are never negative, as detailed  in 
Refs. \cite{gordon07universal, erez08}
and concisely proven in the next section.

{\bf Non-Markovian master equation with non-negative probabilities.}~The non-Markovian master equations (MEs) for an arbitrarily driven (controlled) two-level system (TLS)
presented in \eqref{rateq} have been derived and discussed in Refs. \cite{kofman00acceleration, kofman01universal, kofman04unified, erez08, gordon09cooling} and experimentally verified in Refs. \cite{alvarez10zeno, almog11direct}.
These MEs involve the time-dependent relaxation rates $R_0(t)$ and $R_1(t)$ which can take negative values, since the quantities ${\cal I}_j$, \eqref{Iazd}, are convolutions of a positive spectral 
response function $G_j(\nu)$ with a sinc function which takes positive or negative values.
As a result, the solutions of the MEs for the populations (probabilities) $p_0(t)$ and $p_1(t)$ of the TLS levels are not guaranteed to be non-negative, i.e., to satisfy
\ba
0\le p_k(t)\le1\quad(k=0,1).
\label{nnp1}
\ea
Below we show that the inequalities \eqref{nnp1} hold, at least, up to second order in the system-bath coupling strength.
This means that for a weak coupling, violations of \eqref{nnp1} (if any) are negligibly small.

First, we note that at sufficiently long times, $t\gg\tau_{\rm B}$, the MEs become Markovian and coincide with the Lindblad equation.
In this case, the rates are constant and positive, $R_0,R_1\ge0$, as follows from \eqref{Imkv}. 
The inequalities \eqref{nnp1} are now known to hold.
Generally, the MEs are valid if the couplings of the TLS with the baths are sufficiently weak, so that
\ba
R_0\tau_{\rm B}\ll1,\quad R_1\tau_{\rm B}\ll1.
\label{nnp10}
\ea

Consider now the short times, $t\alt\tau_{\rm B}$, where the non-Markovian effects are important.
Since $p_0(t)+p_1(t)=1$, we rewrite
\ba
p_0(t)=\frac{1-w(t)}{2},\quad p_1(t)=\frac{1+w(t)}{2},
\label{nnp2}
\ea
where $w(t)=p_1(t)-p_0(t)$ is the TLS population inversion.
In terms of $w(t)$, inequalities \eqref{nnp1} are equivalent to
\ba
-1\le w(t)\le1,
\label{nnp5}
\ea
which we now prove.

The condition \eqref{nnp10} implies that at times $\alt\tau_{\rm B}$, the relaxation can be approximated to first order in the relaxation rates.
In this approximation, \eqref{rateq} yields
\ba
w(t)=w(0)[1-J_+(t)]+J_-(t),
\label{nnp11}
\ea
where
\ba
J_\pm(t)=J_0(t)\pm J_1(t)
\label{nnp12}
\ea
and
\ba
J_k(t)=\int_0^td\tau R_k(\tau)\quad\quad(k=0,1).
\label{nnp13}
\ea
From \eqref{Iazd} and \eqref{rateq}, one can check that
\ba
J_0(t),J_1(t)\ge0.
\label{nnp14}
\ea
From \eqref{nnp11} we obtain
\ba
&|w(t)|&\le|w(0)|[1-J_+(t)]+|J_-(t)|\nonumber\\
&&\le1-J_+(t)+J_+(t)=1,
\label{nnp15}
\ea
yielding \eqref{nnp5}.
The second inequality in \eqref{nnp15} follows from the assumption $|w(0)|\le1$ and the relation $|J_-(t)|\le J_+(t)$, resulting from \eqref{nnp12} and \eqref{nnp14}.

{\bf Steady states in the anti-Zeno dynamics regime.}~Now we study the regimes which allow us to operate the setup with a  time-independent steady state $\rho_{\rm ss}$ even inside the AZD regime.
We note that for $t \to \infty$, $I_j\left(\omega_q, t \right)$ reduces to the time-independent form $\pi G_j(\omega_q)$,
thus leading us to the 
Eq. \eqref{wstate}. 
\begin{figure}[h]
\begin{center}
\includegraphics[width= 0.75\columnwidth, angle = 0]{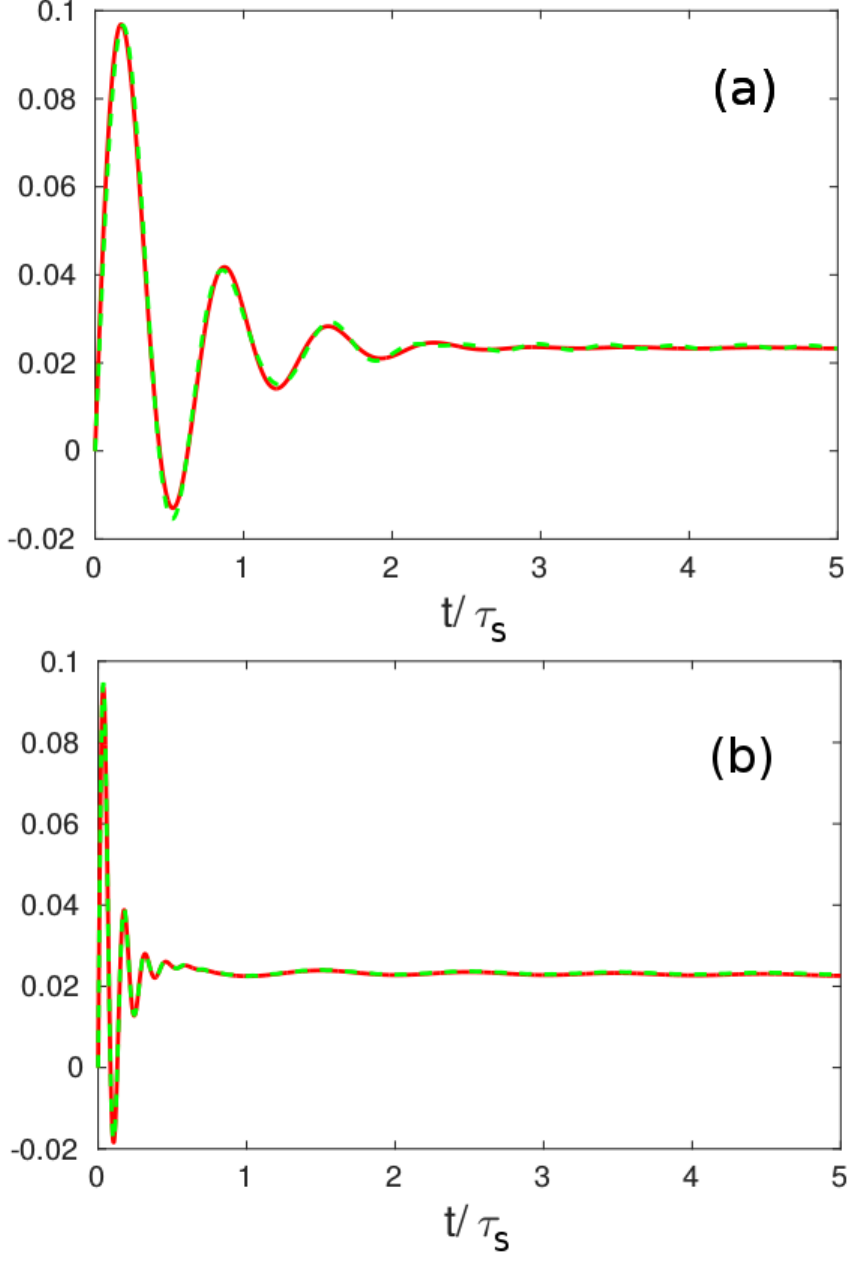}
\end{center}
\caption {{\bf Dissipation rates:} The time dependent rates $\mathcal{I}_{\rm h}(\omega_0 + \Delta_{\rm S}, t)$ (magenta solid line) and 
$\mathcal{I}_{\rm c}(\omega_0 - \Delta_{\rm S}, t)$ (turquoise dashed line) as a function of time-periods $t/\tau_{\rm S}$ for spectral 
functions Eq. (54)
with $\delta = 4.5$ for (a) $\tau_{\rm S} = 2\pi/\Delta_{\rm S} =  2$ and 
(b) $\tau_{\rm S}  = 2\pi/\Delta_{\rm S} = 10$. 
As explained in Methods ``Steady states in the anti-Zeno
dynamics regime'', symmetric spectral functions $G_{\rm h}(\omega + x) = G_{\rm c}(\omega_0 - x)$ for real $x$ lead to 
$\mathcal{I}_{\rm h}(\omega_0 + \Delta_{\rm S}, t) = \mathcal{I}_{\rm c}(\omega_0 - \Delta_{\rm S}, t)$. 
A fast modulation ($\tau_{\rm S} = 2$) results in oscillating and time-dependent $\mathcal{I}_{\rm h, c}(t)$ for large $t/\tau_{\rm S}$, while slow
modulation ($\tau_{\rm S} = 10$) leads to an approximately constant $\mathcal{I}_{\rm h, c}$ even for small $t/\tau_{\rm S}$
Here $\omega_0 = 10, \gamma_0 = 1, \Gamma_{\rm B} = 1$.}
\label{GhGcnm}
\end{figure}
On the other hand, for $t \sim n\tau_{\rm S} \lesssim \tau_{\rm B}$,
$I_j\left(\omega_q, t \right)$ includes contributions from $G_j(\omega_q +\nu)$, where 
\ba
|\nu| \lesssim 1/t = 1/\left(n\tau_{\rm S} \right).
\ea
Further, we consider $\omega_0$, $T_{\rm c}, T_{\rm h}$, and $\Delta_{\rm S} < \omega_0$ large enough, such that $1/t \ll \omega_0 \pm \Delta_{\rm S}, T_{\rm c}, T_{\rm h}$. 
Therefore in this limit the KMS condition gives us
\ba
G_j\left(-\left(\omega_q  + \nu\right)\right) &\approx& e^{-\left(\omega_q + \nu\right)\beta_j}G_j\left(\omega_q  + \nu\right)\non\\
&\approx& e^{-\omega_q \beta_j}G\left(\omega_q + \nu\right).
\ea
This immediately leads us to 
\ba
\mathcal{I}_j(-\omega_q, t) &\approx& e^{-\omega_q \beta_j} \mathcal{I}_j(\omega_q, t),
\ea
and consequently (see Eq. \eqref{wstate})
\ba
w &\approx& \frac{e^{-\left(\omega_0 + \Delta_{\rm S}\right) \beta_{\rm h}}\mathcal{I}_{\rm h}\left(\omega_0 + \Delta_{\rm S},t\right)}{\mathcal{I}_{\rm h}\left(\omega_0 + \Delta_{\rm S},t\right) + \mathcal{I}_{\rm c}\left(\omega_0 - \Delta_{\rm S},t\right)} \non\\
&+&  \frac{e^{-\left(\omega_0 - \Delta_{\rm S}\right)\beta_{\rm c}}\mathcal{I}_{\rm c}\left(\omega_0 - \Delta_{\rm S},t\right)}{\mathcal{I}_{\rm h}\left(\omega_0 + \Delta_{\rm S},t\right) + \mathcal{I}_{\rm c}\left(\omega_0 - \Delta_{\rm S},t\right)},
\ea
where we have considered the two sidebands $q = 1, -1$ only.

The condition
\ba
\mathcal{I}_{\rm h}\left(\omega_0 + \Delta_{\rm S},t\right) \approx \alpha \mathcal{I}_{\rm c}\left(\omega_0 - \Delta_{\rm S},t\right),
\ea
which holds for mutually symmetric bath spectral functions up to a multiplicative factor $G_{\rm h}(\omega_0 + x) \approx \alpha G_{\rm c}(\omega_0 - x)$ for any real $x$ and
positive $\alpha$ (see Fig. \ref{GhGcnm}),
leads to the time-independent steady state $\rho_{\rm ss}$ with (see Eq. \eqref{wstate}) 
\ba
 w \approx \frac{\alpha e^{-\left(\omega_0 + \Delta_{\rm S}\right) \beta_{\rm h}} +  e^{-\left(\omega_0 - \Delta_{\rm S}\right) \beta_{\rm c}}}{\alpha + 1}.
 \label{stdstt}
\ea

{\bf Efficiency and coefficient of performance.}~ The efficiency in the heat engine regime is given by 
\begin{widetext}
\[
\eta = \frac{\oint_{\tau_{\rm C}} \left[J_{\rm h}(t) + J_{\rm c}(t)\right] dt}{\oint_{\tau_{\rm C}} J_{\rm h}(t) dt}
= \frac{\left(\omega_0 + \Delta_{\rm S}\right)\zeta_{\rm h}\oint_{\tau_{\rm C}} \mathcal{I}_{\rm h}(\omega_0 + \Delta_{\rm S}, t) dt + \left(\omega_0 - \Delta_{\rm S}\right)\zeta_{\rm c}\oint_{\tau_{\rm C}} \mathcal{I}_{\rm c}(\omega_0 - \Delta_{\rm S}, t) dt}{\left(\omega_0 + \Delta_{\rm S}\right)\zeta_{\rm h}\oint_{\tau_{\rm C}} \mathcal{I}_{\rm h}(\omega_0 + \Delta_{\rm S}, t) dt},
\]
\end{widetext}
while the coefficient of performance in the refrigerator regime takes the form
\begin{widetext}
\[
{\rm COP} = \frac{\oint_{\tau_{\rm C}} J_{\rm c}(t) dt}{{\oint_{\tau_{\rm C}} \left[J_{\rm h}(t) + J_{\rm c}(t)\right] dt}(t)}
= \frac{\left(\omega_0 - \Delta_{\rm S}\right)\zeta_{\rm c}\oint_{\tau_{\rm C}} \mathcal{I}_{\rm c}(\omega_0 - \Delta_{\rm S}, t) dt}{\left(\omega_0 + \Delta_{\rm S}\right)\zeta_{\rm h}\oint_{\tau_{\rm C}} \mathcal{I}_{\rm h}(\omega_0 + \Delta_{\rm S}, t) dt + \left(\omega_0 - \Delta_{\rm S}\right)\zeta_{\rm c}\oint_{\tau_{\rm C}} \mathcal{I}_{\rm c}(\omega_0 - \Delta_{\rm S}, t) dt},
\]
\end{widetext}
where we have defined 
\ba
\zeta_{\rm h} =  \frac{e^{-(\omega_0 + \Delta_{\rm S})\beta_{\rm h}} - w}{w + 1}, \non\\
\zeta_{\rm c} =  \frac{e^{-(\omega_0 - \Delta_{\rm S})\beta_{\rm c}} - w}{w + 1}.
\ea
One can get the results of the Markovian ($\tau_{\rm C} \to \infty$) limit by replacing $\mathcal{I}_j(-\omega_q, t)$ by 
$G_j\left(\omega_q \right)$. 

Let us consider the integral:
\ba
\mathcal{I}_{\rm h}(\omega_0 + \Delta_{\rm S}, t) &=&  \int^{\infty}_{-\infty}  G_{\rm h}(\nu)   \frac{\sin\left(\left[\nu - \left(\omega_0 + \Delta_{\rm S} \right)\right]t\right)}{\nu - \left(\omega_0 + \Delta_{\rm S} \right)} d\nu \non\\
&\approx& \int_{-\Delta_{\rm S}}^{\infty} G_{\rm h}(\omega_0 + \Delta_{\rm S} + x) \frac{\sin \left(xt \right)}{x} dx, 
\label{iheff}
\ea
where we have defined the variable $x = \nu - \left(\omega_0 + \Delta_{\rm S} \right)$, and taken into account that $G_{\rm h}(\nu) = 0$ for $0 < \nu \leq \omega_0$ (see Eq. \eqref{specsep}), and 
$\sin \left(xt \right)/x$ is small for 
large $|x|$. 

Similarly, we have 
\ba
\mathcal{I}_{\rm c}(\omega_0 - \Delta_{\rm S}, t) &=&  \int^{\infty}_{-\infty}  G_{\rm c}(\nu)   \frac{\sin\left(\left[\nu - \left(\omega_0 - \Delta_{\rm S} \right)\right]t\right)}{\nu - \left(\omega_0 - \Delta_{\rm S} \right)} d\nu \non\\
&\approx& \int_{-\Delta_{\rm S}}^{\infty} G_{\rm c}(\omega_0 - \Delta_{\rm S} - y) \frac{\sin \left(yt \right)}{y} dy, 
\label{iceff}
\ea
where $y = \left(\omega_0 - \Delta_{\rm S} - \nu \right)$, and we have taken into account that $G_{\rm c}(\nu) = 0$ for $\nu \geq \omega_0$ (see Eq. \eqref{specsep}), and 
$\sin \left(yt \right)/y$ is small for 
large $|y|$.

Clearly, for bath spectral functions related by Eq. \eqref{symm}, we have $\mathcal{I}_{\rm h}(-\omega_0 - \Delta_{\rm S}, t) \approx \alpha \mathcal{I}_{\rm c}(-\omega_0 + \Delta_{\rm S}, t)$, which in 
turn results in the efficiency and the coefficient of performance in the non-Markovian anti-Zeno dynamics regime being approximately equal to those in the Markovian dynamics regime (see Fig. \ref{effcop}).

{\bf Quasi-Lorentzian bath spectral functions.}~We focus on  baths characterized by the spectral functions:
\ba 
G_{\rm h}(\nu \geq 0) &=& \frac{1}{N}\sum_{r = 1}^N\left[c_r\frac{\alpha\gamma_0\Gamma_{{\rm B},r}^2\Theta(\nu - \omega_0 - \epsilon)}{(\omega_0  + \Delta_{\rm S} + \delta_r - \nu)^2 + \Gamma_{\rm B, r}^2}\right],\non\\
G_{\rm c}(\nu \geq 0) &=& \frac{1}{N}\sum_{r = 1}^N\left[c_r\frac{\gamma_0\Gamma_{{\rm B},r}^2\Theta(\omega_0 -\epsilon - \nu)\Theta(\nu - \epsilon)}{(\omega_0 - \Delta_{\rm S} - \delta_r - \nu)^2 + \Gamma_{\rm B, r}^2}\right] \non\\
G_{\rm h, c}(-\nu) &=& G_{\rm h, c}(\nu)e^{-\nu \beta_{\rm h, c}},
\label{ghgc3pks}
\ea
where we have considered the KMS condition, $\Theta$ is the step function, $N \in \mathbb{Z}, ~ N > 0$
denotes the number of peaks and
$\Gamma_{{\rm B}, r} = 1/\tau_{{\rm B}, r}> 0$ is the width of the $r$-th peak.
$\delta_r$  are the (real) Lamb self energy shifts, such that
$G_{\rm h}$ ($G_{\rm c}$) is peaked at $\nu = \omega_0 + \Delta + \delta_r$ ($\nu = \omega_0 - \Delta - \delta_r$). 
\begin{figure}[h]
\begin{center}
\includegraphics[width= 0.67\columnwidth, angle = 0]{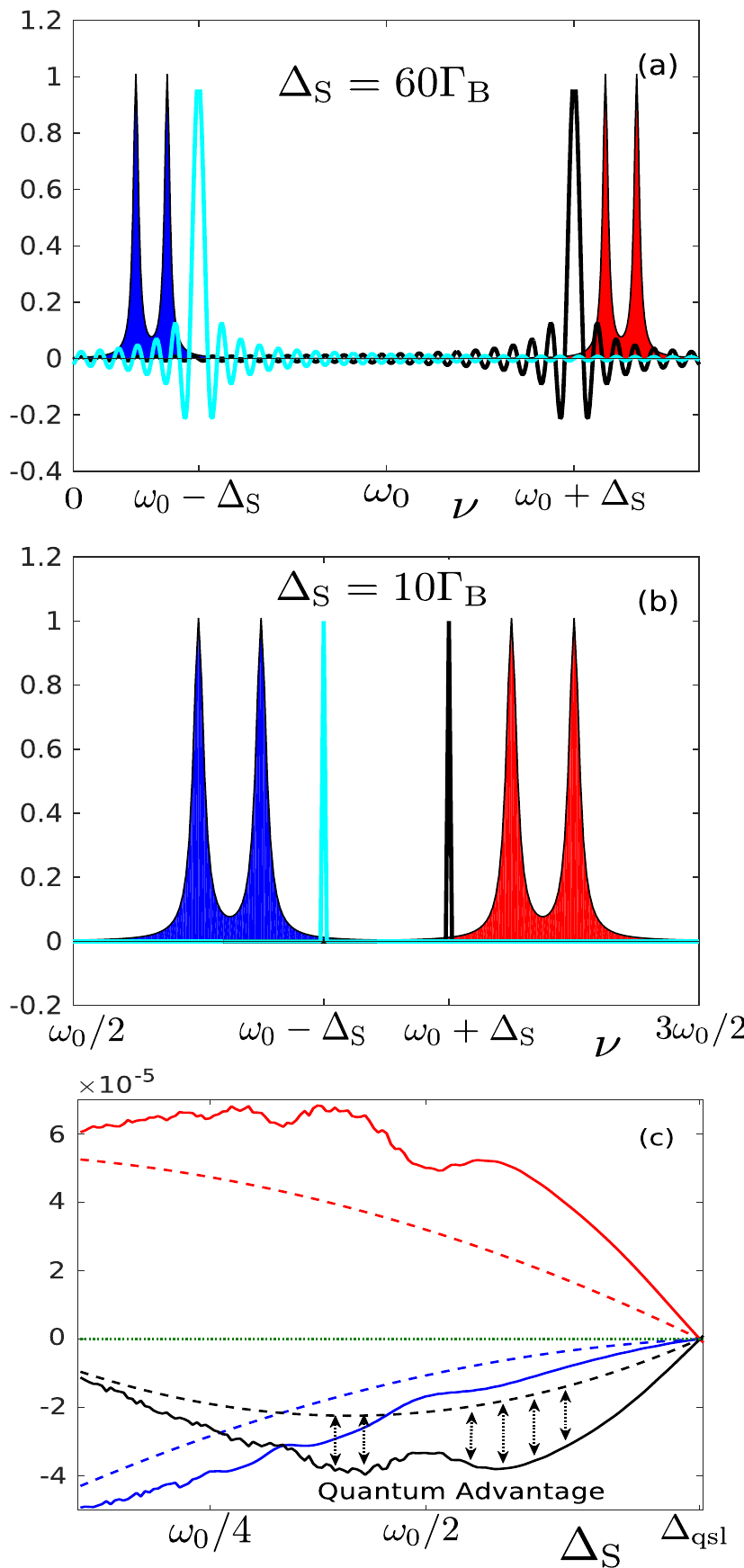}
\end{center}
\caption {{\bf Quantum advantage with double-peaked spectral functions:} Double-peaked spectral functions $G_{\rm h}(\nu)$ (red filled curve) and $G_{\rm c}(\nu)$ (blue filled curve), and the sinc functions
$ {\rm sinc} \left[\left(\nu - \omega_0 - \Delta \right)t\right]$ (black solid curve) and ${\rm sinc} \left[\left(\nu - \omega_0 + \Delta \right)t\right]$ 
(blue solid curve) for (a) fast modulation
$\Delta_{\rm S} = 60\Gamma_{\rm B}$ and (b) slow modulation $\Delta_{\rm S} = 10 \Gamma_{\rm B}$ at $t = 10\tau_{\rm S}$. Fast (slow) modulation results in broadening (narrowing) of the sinc functions, 
thus leading to enhanced (reduced) overlap with the spectral functions. (c) Power $\overline{\dot{W}}$ (black lines) and heat currents
$\overline{J_{\rm h}}$ (red lines) and  $\overline{J_{\rm c}}$ (blue lines)
averaged over $n = 10$ modulation periods (solid lines) and the same obtained under the Markovian approximation for
long cycles ($n \to \infty$)  (dashed lines), versus the modulation frequency $\Delta_{\rm S}$.
A significant quantum advantage (shown by dotted double-arrow lines) is obtained for 
fast modulation,  when broadening of the sinc functions yields an output power boost by a factor greater than $2$ in the heat engine regime.
The green dotted line corresponds to zero power and currents.  Here $\lambda = 0.2, \omega_0 = 20, N = 2, \delta_{1} = 2, \delta_{2} = 4, \alpha = 1, \gamma_0 = 1, \Gamma_{\rm B} = 0.2, \epsilon = 0.01, \beta_{\rm h} = 0.0005, \beta_{\rm c} = 0.005$.}
\label{sinc3pks}
\end{figure}

As seen from Eq. \eqref{ghgc3pks}, we consider bath spectral functions with different resonance frequencies
($= \omega_0  \pm \Delta_{\rm S} \pm \delta_r $) for different modulation rates $\Delta_{\rm S}$. As mentioned above, this ensures that the detuning between the $r$-th resonance frequency of a
bath spectral function, 
and the maximum of the corresponding sinc function, is always $\delta_{r}$, and is independent of the modulate rate $\Delta_{\rm S}$. For example, this can be implemented by choosing different baths 
for operating thermal machines with different modulation frequencies. Consequently, any enhancement in 
heat currents and power originate from the broadening of the sinc functions, rather than from the shift of the maxima of the sinc functions.
Here $c_r \geq 0$ is the weight of the $r$-th term in the sums in Eq. \eqref{ghgc3pks}. A non-zero (but small) $\epsilon > 0$ ensures that 
$G_{\rm c}(\nu)$ and $G_{\rm h}(\nu)$ vanish at $\nu = 0$, thus resulting in vanishing thermal excitations and entropy at the absolute zero temperature,  as is demanded by the third law of thermodynamics \cite{kolar12, masanes17a}. Since 
$G_{\rm c}(\nu = \omega_0) = G_{\rm c}(\nu = \omega_0) = 0$,  the $0$-th sideband ($q = 0$) does not contribute to the dynamics.
Figure \ref{sinc3pks} shows the quantum advantage obtained for bath spectral functions of the form Eq. \eqref{ghgc3pks} with $N = 2$ (double-peaked functions).

For the single-peaked case ($N = 1$), the above functions Eq. \eqref{ghgc3pks} reduce to quasi-Lorentzian spectral functions of the form 
\ba
G_{\rm h}(\nu \geq 0) &=& \frac{\alpha\gamma_0\Gamma_{\rm B}^2 \Theta(\nu - \omega_0 - \epsilon)}{\left(\omega_0 + \Delta_{\rm S} + \delta - \nu\right)^2 + \Gamma_{\rm B}^2},\non\\
G_{\rm c}(\nu \geq 0) &=& \frac{\gamma_0\Gamma_{\rm B}^2 \Theta(\omega_0 - \epsilon - \nu )\Theta(\nu - \epsilon)}{\left(\omega_0 -  \Delta_{\rm S} - \delta - \nu\right)^2 + \Gamma_{\rm B}^2}, \non\\
G_{\rm h, c}(-\nu) &=& G_{\rm h, c}e^{-\nu \beta_{\rm h, c}}.
\label{ghgceq}
\ea
The condition $\delta = 0$ results in the spectral functions and the sinc function attaining maxima
 at the same frequencies, viz., at $\nu = \omega_0 \pm \Delta_{\rm S}$.

{\bf Super-Ohmic bath spectral functions.}~We also consider super-Ohmic bath spectral functions of the form 
\ba
G_{\rm h}(\nu \geq 0) &=& \Theta\left(\nu -  \nu_{\rm h}\right) \alpha \gamma_0 \frac{\left(\nu - \nu_{\rm h}\right)^s}{\bar{\nu}^{s-1}}e^{\left[-\left(\nu - \nu_{\rm h} \right)/\bar{\nu}\right]} \non\\
G_{\rm c}(\nu \geq 0) &=& \Theta\left(\nu_{\rm c} -  \nu\right)\Theta\left(\nu - \epsilon\right) \gamma_0 \frac{\left(\nu_{\rm c} - \nu\right)^s}{\bar{\nu}^{s-1}}e^{\left[-\left(\nu_{\rm c} - \nu \right)/\bar{\nu} \right]}\non\\
G_{\rm h, c}(-\nu) &=& G_{\rm h, c}(\nu)e^{-\nu \beta_{\rm h, c}},
\label{specso}
\ea
with the origin shifted from $\nu = 0$ by 
\ba
\nu_{\rm h} = \omega_0 + \Delta_{\rm S} - \delta \non\\
\nu_{\rm c} = \omega_0 - \Delta_{\rm S} + \delta
\ea
Here $s > 1$, and 
\ba
0 < \delta \ll \Delta_{\rm S}, \omega_0, \omega_0 - \Delta_{\rm S}
\ea
ensures that $G_{\rm h, c}(\nu)$ is non-zero at the 
maxima of the sinc functions at $\omega_0 \pm \Delta_{\rm S}$. As before, a small $\epsilon > 0$ guarantees that $G_{\rm c}(\nu = 0) = 0$, and we consider $\Delta_{\rm S}$-dependent $\nu_{\rm h}$ and $\nu_{\rm c}$, to ensure that any enhancement in heat currents and power are due to the broadening of 
the sinc functions for fast modulations, rather than due to the shifting of the peaks of the sinc functions. 

{\bf Thermal machines with arbitrary (asymmetric) spectral functions.}~ We consider 
\ba
G_{\rm h}(\omega_0 + \nu) = \alpha G_{\rm c}(\omega_0 - \nu) + \tilde{\chi}(\nu),
\label{nonsymm}
\ea
where, as before, $\alpha > 0$ and $\tilde{\chi}(\nu)$ is an arbitrary real function of $\nu$. We then have 
\ba
\mathcal{I}_{\rm h}(\omega_0 + \Delta_{\rm S}) = \alpha\mathcal{I}_{\rm c}(\omega_0 - \Delta_{\rm S}) + \chi(t), \non\\
\chi(t) = \int_{-\Delta_{\rm S}}^{\infty} \tilde{\chi}(\nu) \frac{\sin\left(\nu t \right)}{\nu} d\nu.
\ea
and 
\ba
\dot{\chi}(t) = \int_{-\Delta_{\rm S}}^{\infty} \tilde{\chi}(\nu) \cos\left(\nu t \right) d\nu.
\ea
In this case, we get a time-dependent steady-state with 
\ba
w(t) &\approx& \frac{e^{-\left(\omega_0 + \Delta_{\rm S}\right) \beta_{\rm h}}\left[\alpha\mathcal{I}_{\rm c}\left(\omega_0 - \Delta_{\rm S},t\right) + \chi(t)\right]}{\left(\alpha + 1 \right)\mathcal{I}_{\rm c}\left(\omega_0 - \Delta_{\rm S},t\right) + \chi(t)} \non\\
&+&  \frac{e^{-\left(\omega_0 - \Delta_{\rm S}\right)\beta_{\rm c}}\mathcal{I}_{\rm c}\left(\omega_0 - \Delta_{\rm S},t\right)}{\left(\alpha + 1 \right)\mathcal{I}_{\rm c}\left(\omega_0 - \Delta_{\rm S},t\right) + \chi(t)},
\ea
Therefore, the rate of change of $w(t)$ with time is given by 
\begin{widetext}
\[
\dot{w}(t) = \frac{\left[\chi(t)\dot{\mathcal{I}}_{\rm c}\left(\omega_0 - \Delta_{\rm S},t\right) - \dot{\chi}(t)\mathcal{I}_{\rm c}\left(\omega_0 - \Delta_{\rm S},t\right)\right]\left[\left(e^{-\left(\omega_0 + \Delta \right)\beta_{\rm h}} - e^{-\left(\omega_0 - \Delta \right)\beta_{\rm c}} \right) - \left(\alpha + 1 \right)e^{-\left(\omega_0 + \Delta \right)\beta_{\rm h}}\right]}{\left[\left(\alpha + 1 \right)\mathcal{I}_{\rm c}\left(\omega_0 - \Delta_{\rm S},t\right) + \chi(t) \right]^2}
\]
\end{widetext}
One can still operate the setup as a cyclic thermal machine for a time $t \leq \tilde{t}$, as long as $\chi(t)$ and $\dot{\chi}(t)$ are small enough so as to ensure
\ba
\dot{w}_{\rm max} \ll \tilde{t}^{-1},
\ea
where $\dot{w}_{\rm max}$ is the maximum value attained by $|\dot{w}(t)|$ in the time interval $0 \leq t \leq \tilde{t}$.

\section*{Data availability}
All relevant data are available to any reader upon reasonable request.

\section*{Code availability}
All relevant codes are available to any reader upon reasonable request.

\section*{Acknowledgements}
We acknowledge the support of ISF, DFG (FOR 7024), EU (PATHOS,  FET Open), QUANTERA (PACE-IN)  and VATAT.

\section*{AUTHOR CONTRIBUTIONS}
G.K. and V.M. conceived the idea. V.M. and A.G.K. performed the
analytical calculations. V.M. did the numerical simulations. All
authors contributed to the interpretations of the results and
to the writing of the manuscript.

\appendix

\section{Quantum refrigeration above the quantum speed limit}
\label{app1}

The quantum speed limit $\Delta_{\rm qsl}$ is defined as the largest modulation rate which 
allows the system to operate as a heat engine, i.e., a modulation with
$\Delta_{\rm S} > \Delta_{\rm qsl}$ results in $\overline{\dot{W}} > 0$ \cite{mukherjee16speed, deffner17quantum}.
Above this modulation rate, the setup stops acting as a heat engine, and instead starts operating as a refrigerator, where the heat current flows from the cold bath to the hot bath, in 
presence of  $\overline{\dot{W}} > 0$ (see Fig. \ref{refr}) \cite{klimovsky13minimal, klimovsky15, mukherjee16speed}.
Analysis of the heat currents (see Eq. \eqref{Jhc}) yields, for bath temperatures $T_h > T_c$,
\ba
\Delta_{\rm qsl} =  \frac{2\pi}{\tau_{\rm qsl}} = \omega_0 \frac{T_h - T_c}{T_h + T_c}.
\ea
Interestingly, the same result is obtained for Markovian heat engines characterized by $\tau_{\rm S} \gg \tau_{\rm B}$. Therefore, under fast modulation, 
the quantum thermal machine operates as a quantum refrigerator  for 
\ba
\tau_{\rm S} < \tau_{\rm qsl} \ll \tau_{\rm B}
\ea
and as a  heat engine for
\ba
\tau_{\rm qsl} < \tau_{\rm S} \ll \tau_{\rm B}.
\ea

\section{Zeno dynamics}
\label{app2}

In contrast to the advantageous AZD, ultra-fast modulations with $\tau_{\rm B}, ~\delta^{-1} \gg \tau_{\rm C}$,
lead to the Zeno regime,
where the maximum of 
$t {\rm sinc}\left(\nu t \right)$ at $\nu \to 0$ and $t = n\tau_{\rm S} = \tau_{\rm C}$ is given by
\ba
\lim_{\nu \to 0} t {\rm sinc}\left(\nu t \right) \to \tau_{\rm C} = n\tau_{\rm S} \to 0.
\ea
Consequently, the
convolutions $\mathcal{I_{\rm c, h}}(\omega_q,t) \to 0$, resulting in vanishing heat currents and power (see Figs. \ref{regimes} and \ref{zazdreg}). However, the Zeno regime might 
be beneficial for work 
extraction in the presence of appreciable system-bath correlations, that are neglected here \cite{klimovsky13work}.

\section{Energetic cost}
\label{app3}

One can estimate the energetic cost $\mathcal{E}_{\rm dec}$ ($\mathcal{E}_{\rm rec}$) of decoupling (recoupling) the WF and the
baths by changing $\omega(t)$ from $\omega_0$ ($\tilde{\omega}$) to 
$\tilde{\omega}$ ($\omega_0$) as 
\ba
\mathcal{E}_{\rm dec} &=& \int_{\delta t_{\rm dec}}  {\rm Tr}\left[\frac{1}{2}\dot{\omega}(t)\hat{\sigma}_z\rho_{ss}\right] dt \non\\
&=&  -\int_{\delta t_{\rm rec}} {\rm Tr}\left[\frac{1}{2}\dot{\omega}(t)\hat{\sigma}_z\rho_{ss}\right] dt =  -\mathcal{E}_{\rm rec}, 
\ea
where $\delta t_{\rm dec}$ and $\delta t_{\rm rec}$ are respectively the short time intervals during which the WF-bath interactions are effectively being turned off, 
and turned on, respectively, due to changing $\omega(t)$ \cite{alicki79the}.
Consequently, for the cyclic thermal machine described here, the total cost of decoupling and recoupling the WF with the baths is $\mathcal{E}_{\rm dec} + \mathcal{E}_{\rm rec} = 0$.

\renewcommand\bibsection{\section*{\refname}}
\renewcommand\refname{References}
\bibliographystyle{naturemag}
%\bibliography{Fast_QHE}

\end{document}